\documentclass{IEEEoj}
\usepackage{cite}
\usepackage{amsmath,amssymb,amsfonts}
\usepackage{graphicx,color}
\usepackage{xcolor}
\usepackage{textcomp}
\usepackage[normalem]{ulem}
\usepackage[linesnumbered,ruled,vlined]{algorithm2e}
\usepackage{float}
\usepackage{tabularx}
\usepackage{threeparttable}
\usepackage{caption}
\usepackage{booktabs}
\usepackage{multirow}
\usepackage{makecell}
\usepackage{subcaption}
\usepackage{hyperref}

\def\BibTeX{{\rm B\kern-.05em{\sc i\kern-.025em b}\kern-.08em
    T\kern-.1667em\lower.7ex\hbox{E}\kern-.125emX}}
\AtBeginDocument{\definecolor{ojcolor}{cmyk}{0.93,0.59,0.15,0.02}}
\def\OJlogo{\vspace{-4pt}\hskip-4pt\includegraphics[height=18pt]{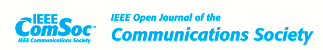}}
\begin{document}
\receiveddate{XX Month, XXXX}
\reviseddate{XX Month, XXXX}
\accepteddate{XX Month, XXXX}
\publisheddate{XX Month, XXXX}
\currentdate{11 January, 2024}
\doiinfo{OJCOMS.2024.011100}

\title{Context-Aware Predictive Coding: A Representation Learning Framework for WiFi Sensing}

\author{Borna Barahimi\IEEEauthorrefmark{1} \IEEEmembership{(Student Member, IEEE)}, Hina Tabassum\IEEEauthorrefmark{1} \IEEEmembership{(Senior Member, IEEE)}, Mohammad Omer\IEEEauthorrefmark{2}, AND Omer Waqar\IEEEmembership{(Senior Member, IEEE)} \IEEEauthorrefmark{1, 3}}
\affil{Department of Electrical Engineering and Computer Science at York University, Toronto, ON. M3J 1P3 Canada}
\affil{Cognitive Systems Corp., Waterloo, ON. N2L 0A9 Canada}
\affil{School of Computing, University of the Fraser Valley, BC. V2S 7M8 Canada}
\corresp{CORRESPONDING AUTHOR: Borna Barahimi (e-mail: bornab@yorku.ca).}
\authornote{This research was supported by an Alliance Grant funded by the Natural Sciences and Engineering Research Council of Canada (NSERC).}
\markboth{Preparation of Papers for IEEE OPEN JOURNALS}{Author \textit{et al.}}

\begin{abstract}
WiFi sensing is an emerging technology that utilizes wireless signals for various sensing applications. However, the reliance on supervised learning and the scarcity of labelled data and the incomprehensible channel state information (CSI) data pose significant challenges. These issues affect deep learning models' performance and generalization across different environments. Consequently, self-supervised learning (SSL) is emerging as a promising strategy to extract meaningful data representations with minimal reliance on labelled samples.
In this paper, we introduce a novel SSL framework called Context-Aware Predictive Coding (CAPC), which effectively learns from unlabelled data and adapts to diverse environments. CAPC integrates elements of Contrastive Predictive Coding (CPC) and the augmentation-based SSL method, Barlow Twins, promoting temporal and contextual consistency in data representations. This hybrid approach captures essential temporal information in CSI, crucial for tasks like human activity recognition (HAR), and ensures robustness against data distortions. Additionally, we propose a unique augmentation, employing both uplink and downlink CSI to isolate free space propagation effects and minimize the impact of electronic distortions of the transceiver.
Our evaluations demonstrate that CAPC not only outperforms other SSL methods and supervised approaches, but also achieves superior generalization capabilities. Specifically, CAPC requires fewer labelled samples while significantly outperforming supervised learning by an average margin of 30.53\% and surpassing SSL baselines by 6.5\% on average in low-labelled data scenarios. Furthermore, our transfer learning studies on an unseen dataset with a different HAR task and environment showcase an accuracy improvement of 1.8\% over other SSL baselines and 24.7\% over supervised learning, emphasizing its exceptional cross-domain adaptability. These results mark a significant breakthrough in SSL applications for WiFi sensing, highlighting CAPC’s environmental adaptability and reduced dependency on labelled data.
\end{abstract}


\maketitle

\section{Introduction}
\raggedbottom
\IEEEPARstart{T}{he} ubiquity of the internet has led to a significant increase in WiFi-enabled devices and access points (APs) within various environments, such as commercial and residential areas. WiFi APs have advanced from being simple routers to functioning as sensor devices for human sensing applications, enabled by the analysis of wireless signal characteristics like received signal strength indicator (RSSI) or fine-grained channel state information (CSI). CSI includes wireless channel properties such as amplitude and phase, while RSSI measures the power level of the received signal affected by distance and obstructions.
CSI captures variations in radio frequency (RF) signals as they move through a physical space, interacting with objects or human bodies, causing reflection, diffraction, and scattering. These multi-path effects convey valuable information about the environment, including human movements, locations, and the state of objects \cite{9900419}.  

WiFi sensing holds the promise to redefine the sensing paradigm. WiFi sensing places emphasis on privacy and facilitates ubiquitous, non-invasive sensing as users do not need to carry  sensors \cite{9831898}. WiFi sensing is cost-efficient as it capitalizes on existing WiFi infrastructure. Unlike systems limited by line-of-sight (LOS) constraints, wireless signals provide rich data through reflection and diffraction, even in non-line-of-sight (NLOS) scenarios where obstructions exist between the target and the WiFi device. Notably, these signals can operate in the dark, offering a round-the-clock functionality that cameras cannot match. WiFi signals permit the extraction of specific details, such as human position and vital signs, without visual information.  WiFi sensing enables applications such as localization \cite{8397121}, human activity recognition (HAR) \cite{shi2022environment}, gesture recognition \cite{9233449, pu2013whole, 9366929}, and  respiration detection \cite{10.1145/3351279}.

Traditionally, WiFi sensing relies on signal processing methods like Fresnel zone to model CSI \cite{wang2016human}, Hampel filtering for pre-processing \cite{9831898}, and discrete wavelet transforms \cite{7875148} for signal transformation. However, these methods struggle to capture the nuances of complex human activities, such as gait \cite{yang2023sensefi}. 
Moreover, the methods are not typically reusable, versatile, or scalable for new tasks or environments \cite{10.1145/3310194}.

Recently, learning-based models, especially deep learning, have been pursued due to their enhanced ability to extract  CSI patterns for complex activities \cite{Zhang_Tang_Li_Fang_Nurmi_Wang_2018}. However, the success and adaptability of these models are limited by the scale and diversity of the training samples and labels. The process of data collection and annotation for CSI is particularly demanding and slow, necessitating simultaneous labelling since they are unreadable by humans, in contrast to computer vision datasets where labels can be applied after collection. Although efforts to create extensive public datasets such as SignFi \cite{10.1145/3191755} and Widar \cite{7znf-qp86-20} have been made, these datasets remain substantially smaller than those for computer vision like ImageNet \cite{deng2009imagenet}. Furthermore, most CSI datasets are curated in controlled lab environments and may not perform well in varying real-world scenarios. Insufficient data and/or training labels can lead to failures of learning-based models in unseen settings \cite{yang2022autofi}.

\begin{figure}[h]
    \centering
    \includegraphics[width=1\linewidth]{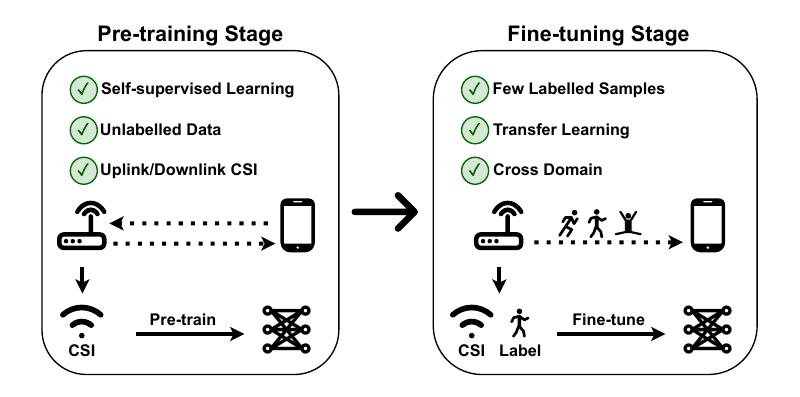}
    \caption{Illustration of the proposed CAPC framework. Initially, unlabelled uplink and downlink CSI are utilized to pre-train an encoder through an unsupervised approach. Subsequently, the model undergoes fine-tuning using a limited set of labelled CSI for the HAR task in the unseen environment.}
    \label{fig:abstract}
\end{figure}

To bridge the gap between learning-based models and realistic WiFi sensing, recent methods have focused on models that are data efficient and work with a limited amount of labelled data. More specifically, utilizing unlabelled data to facilitate the model training process is viable in the presence of scarce training labels. Consequently, self-supervised learning (SSL) has gained much attention recently for exploiting unlabelled data to learn powerful representations for deep learning models. Compared with models trained on fully labelled data (i.e., supervised models),
SSL approaches allow models to pre-train on vast amounts of unlabelled data, learning general features and representations that can later be fine-tuned with a smaller set of labelled data, thus enabling the extraction of effective representations and achieving comparable performance when fine-tuned with a small percentage of the labels \cite{chen2020simple}. 
This process involves designing specific pretext tasks, such as predicting missing parts of the input data or identifying transformations
to encourage the model to learn meaningful representations from unlabelled data, thereby circumventing the need for extensive manual annotation. This paradigm shift towards data-efficient approaches is particularly pertinent in the context of WiFi sensing, where labelled CSI data is inherently scarce and challenging to obtain. By harnessing SSL techniques, recent methods strive to meet the practical demands of WiFi sensing applications. These approaches leverage unlabelled data to enhance model training, enabling the development of robust and adaptable WiFi sensing systems that can generalize effectively across diverse environmental conditions \cite{yang2022autofi}, \cite{shen2023bts}, \cite{ji2022sifall}.

To date, various SSL methods have been considered for WiFi sensing \cite{10.1145/3495243.3560529, 9522151, 9939099,10008537, chen2020simple} that utilize a contrastive loss function. This function trains an encoder to produce similar representations for different augmented versions of the same data sample, known as positive samples. Simultaneously, it ensures that these representations are distinct from those of other samples, termed negative samples \cite{chen2020simple}. 

Nonetheless, the aforementioned research works do not consider the temporal aspects of CSI, wireless propagation channel, or transceiver characteristics.  In contrast to images that primarily exhibit spatial features, CSI, as time-series data, is predominantly defined by its temporal relationships \cite{franceschi2020unsupervised},  \cite{10233092}. Moreover, common augmentation methods designed for computer vision tasks, like colour distortion or rotation, typically do not align well with the nature of CSI data \cite{gidaris2018unsupervised}. Finally, the recent developments in non-contrastive methods, which do not rely on negative samples \cite{zbontar2021barlow}, \cite{bardes2022vicreg}, have yet to be explored in the WiFi sensing domain.

In this study, we propose a novel SSL framework tailored for CSI WiFi sensing, named \textbf{C}ontext-\textbf{A}ware \textbf{P}redictive \textbf{C}oding (\textbf{CAPC}). CAPC combines the strengths of two SSL methods—Barlow Twins \cite{zbontar2021barlow}  and Contrastive Predictive Coding (CPC) \cite{oord2018representation}—to learn robust and informative representations from CSI. Barlow Twins is a SSL approach that aims to learn invariant and redundant-free representations by maximizing similarity between augmented views of the same data, while reducing feature redundancy. In contrast, CPC focuses on capturing temporal dependencies by predicting future segments of sequential data from the current context, using contrastive learning to distinguish between positive future samples and negatives.

Inspired by these methods, CAPC utilizes a twin-branch design where each branch processes multiple augmented views of CSI samples, employing WiFi-specific augmentations including a novel one that leverages both uplink and downlink CSI. CAPC leverages CPC's contrastive approach to predict future representations and capture temporal dynamics, while also incorporating Barlow Twins' non-contrastive loss to ensure consistency and redundancy reduction across different augmented views. This hybrid approach enables CAPC to learn temporally informative and contextually consistent representations that are robust to augmentations and distortions, enhancing the model's ability to generalize across different environments and sensing tasks.

Our key contributions can be summarized as follows:

$\bullet$ We develop a novel SSL framework, CAPC, along with considerations for time-series and wireless propagation channel properties. CAPC utilizes a contrastive predictive method to capture temporal dynamics and a non-contrastive method for invariant, contextually consistent representations.

$\bullet$  We introduce a novel augmentation that leverages both uplink and downlink CSI, enhancing the model's generalization capabilities by isolating free space propagation effects and minimizing electronic distortions. 

$\bullet$  We perform a quantitative comparative analysis of several common augmentations for time-series data as well as our proposed one to find the best combination suited for WiFi sensing and different SSL methods.

$\bullet$ We demonstrate that CAPC requires fewer labelled samples while achieving superior performance compared to other SSL baselines, with an average improvement margin of 6.5\%. Additionally, CAPC outperforms supervised learning by an average margin of 30.53\% in low-labelled data scenarios. These results are validated under both linear and semi-supervised evaluations in unseen environments using the SignFi dataset \cite{10.1145/3191755}.

$\bullet$ We evaluate the transfer learning performance of CAPC on the UT HAR dataset \cite{8067693}, where CAPC representations exhibit superior generalization to the new HAR task in an unseen environment. CAPC outperforms SSL baselines by 1.8\% and supervised learning by 24.7\% in accuracy, further showcasing its cross-domain capabilities.

$\bullet$ We have made CAPC’s source code available on GitHub at \href{https://github.com/bornabr/CAPC}{https://github.com/bornabr/CAPC}.

The remainder of the paper is structured as follows: Section~\ref{sec:backgrounds} discusses the WiFi sensing preliminaries and related work. Section~\ref{sec:method} describes the proposed method. Section~\ref{sec:settings} provides details on the experimental setup and evaluation methods. Section~\ref{sec:experiments} presents the numerical results of the experiments, followed by the conclusion in Section~\ref{sec:conclusion}.

\section{WiFi Sensing Preliminaries and Related Work}
\label{sec:backgrounds}

In this section, we will first briefly cover the preliminaries of CSI-based WiFi sensing and then review the existing state-of-the-art for deep learning-based WiFi sensing.

\subsection{Channel State Information}

CSI describes the propagation of wireless signals between a transmitter and a receiver, accounting for the effects of environmental reflections.  CSI provides a detailed characterization of the wireless environment, influenced by variations in the positions and movements of the transmitter, receiver, and nearby objects. It is mathematically represented as:
\begin{equation}
    \boldsymbol{y}_{t,s,a} = \boldsymbol{H}_{t,s,a}\boldsymbol{x}_{t,s,a} + \boldsymbol{\eta}_{t,s,a},
\end{equation}
where $\boldsymbol{H}_{t,s,a}$ is the CSI matrix that transforms the transmitted signal $\boldsymbol{x}_{t,s,a}$ into the received signal $\boldsymbol{y}_{t,s,a}$, for the $t$th packet, $s$th subcarrier, and $a$th transmitter-receiver antenna link. Here,  $\boldsymbol{\eta}_{t,s,a}$ represents the additive white Gaussian noise (AWGN) at the receiver associated with the specific transmitter-receiver link $a$. Human activities can alter the multipath characteristics of the wireless channel, affecting the channel matrix $\boldsymbol{H}$ due to reflection, scattering, and diffraction. The complex values in the matrix $\boldsymbol{H}$ represent the amplitude and phase of the signals with amplitude being more stable and thus preferred for WiFi sensing applications \cite{yang2023sensefi} due to random phase offsets from a single antenna \cite{liu2020human}.

For extracting and recording CSI from WiFi devices, three widely used tools are available: the Intel 5300 NIC \cite{10.1145/1925861.1925870}, the Atheros CSI Tool \cite{10.1145/2789168.2790124}, and the Nexmon CSI Tool \cite{nexmon:project}. These tools allow for capturing wireless signals with varying numbers of subcarriers and bandwidths. 

\subsection{Related Work}
\subsubsection{Supervised Learning for WiFi Sensing}

Recent advances in deep learning (DL)-enabled sensing have demonstrated remarkable potential in extracting and modeling complex patterns from CSI without intensive feature engineering \cite{xiao2019csigan, 10.1145/3310194}.
Methods based on Long Short-Term Memory (LSTM) networks were used to extract temporal activity-related information from CSI data \cite{zhuravchak2022human}, \cite{8514811} and achieved state-of-the-art accuracy at the time. Convolutional Neural Networks (CNNs), commonly used to extract spatial features, have been used to segment CSI based on activities  \cite{9235578} and compress CSI data for cloud-based sensing \cite{barahimi2024rscnet}. Additionally, THAT \cite{li2021two} proposed a two-stream Transformer-based model to utilize both time-over-channel and channel-over-time features and out-performed LSTM and CNN based methods. ResNet architectures with attention mechanisms for gesture recognition \cite{9759238} were proposed as well for cross-domain HAR.
Notably, some studies have also addressed the environmental dependencies of DL methods.  Widar3.0 \cite{9516988} used a hand-crafted environment-independent feature called body-coordinate velocity profile (BVP) as the input of their DL model and AFEE-MatNet \cite{shi2022environment} applied a matching network to learn common features among environments.

\subsubsection{Semi-supervised learning for WiFi Sensing}

Nevertheless, the aforementioned research works follow a supervised approach, thus necessitating large volumes of labelled CSI data.   A fundamental challenge is to train the model without labels, while enabling generalization across different environment settings.
Recently, a handful of research works considered semi-supervised learning to reduce the reliance on labelled datasets. For instance, in WiADG \cite{8487345}, where a pre-trained encoder, initially trained in a supervised manner, is fine-tuned for new environments using adversarial networks. This semi-supervised approach can adapt to new settings without the additional labelled data in the target environment, though it is contingent on having pre-labelled data for the source environment. Fidora \cite{9745151} used a semi-supervised approach by employing Variational Auto Encoders (VAEs) to augment the labelled dataset.  Subsequently, a feature extractor is applied to generate a representation from labelled and unlabelled samples. These representations are then processed through a decoder and classifier for  CSI reconstruction and classification, respectively. SiFall \cite{ji2022sifall} also leveraged VAEs in conjunction with anomaly detection techniques for fall detection. BTS \cite{shen2023bts} proposed a semi-supervised teacher-student learning approach inspired by BYOL \cite{grill2020bootstrap} to solve time-varying effects induced by environment changes.

\subsubsection{Self-supervised learning for WiFi Sensing}

SSL is emerging as a powerful technique that leverages unlabelled data to train deep learning models, generating effective representations without the need for explicit labels. However, it requires knowledge of what makes some samples semantically close to others \cite{balestriero2022contrastive}. In the initial phase, the model undergoes pre-training using unlabelled dataset, thereby generating useful representations from the data without relying on explicit labels. In the subsequent stage, the model shifts to supervised learning using a limited labelled target dataset. During this phase, fine-tuning takes place, exploiting the previously learned representations to improve the performance. 

Recent SSL-based WiFi sensing methods, such as RF-URL\cite{10.1145/3495243.3560529}, Lau et al. \cite{10008537}, STF-CSL\cite{9522151}, DualConFi\cite{9939099}, and CLAR\cite{10599540}, learn invariant representations by contrasting or aligning different views of the input samples, which are created through various augmentation techniques. These methods focus on creating representations that minimize the distance between similar instances in an embedding space.
RF-URL used Doppler-frequency-spectrum, Angle of Arrival, and Time of Flight augmentations as well as InfoNCE contrastive loss function. STF-CSL  integrated Short-Time-Fourier-Neural Networks (STFNets) in their encoders, with a variety of frequency and time-domain augmentations. In \cite{10008537}, different views of CSI corresponding to a specific activity (or sample) captured by several receivers placed at various locations are considered positive samples, while views of other unrelated samples are treated as negative samples. CLAR, a more recent work, utilizes diffusion models to generate augmented samples. Except for RF-URL, these methods adopt a contrastive learning strategy akin to SimCLR \cite{chen2020simple}, using the NT-Xent loss function. More recently, AutoFi\cite{yang2022autofi} has emerged with a non-contrastive geometric SSL method and few-shot learning for WiFi sensing.

However, the aforementioned research overlooks the temporal aspects of CSI, as well as wireless propagation channel, and transceiver characteristics. Additionally, the considered augmentation techniques often do not align well with the inherent nature of CSI data. Furthermore, recent advancements in non-contrastive methods \cite{zbontar2021barlow, bardes2022vicreg}, which eliminate the need for negative samples, have not been explored in the WiFi sensing domain.


\section{Context-Aware Predictive Coding Framework for Self-Supervised WiFi Sensing}
\label{sec:method}

\begin{figure*}
    \centering
    \includegraphics[width=0.90\linewidth]{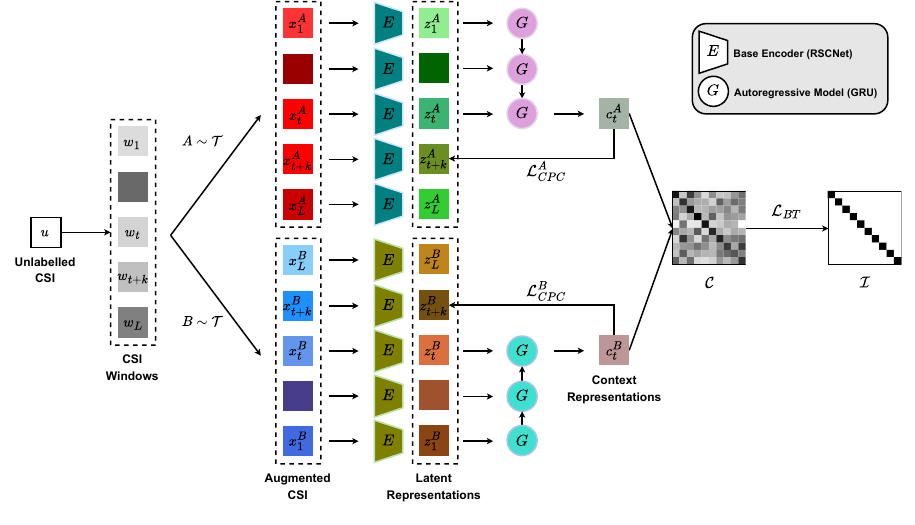}
    \caption{Overview of the CAPC's architecture. Here, $\boldsymbol{w_t}$ denotes a window of sample $\boldsymbol{u}$. The symbols $\boldsymbol{x_t}$, $\boldsymbol{z_t}$, and $\boldsymbol{c_t}$ represent the augmented CSI for window $t$, the latent representation of this window, and the accumulated context embedding up to window $t$, respectively. Different colours signify distinction in the windows, their representations, and model parameters between branches A and B.}
    \label{fig:ssl-framework}
\end{figure*}


In this section, we describe the components of our novel CAPC framework designed to improve CSI-based WiFi sensing as well as the novel model-based augmentation.
CAPC involves pre-training an encoder with unlabelled CSI data to generate feature-rich representations suitable for downstream WiFi sensing applications. 

\subsection{Overview}

The components of our proposed CAPC framework are as follows:

$\bullet$ \textbf{CSI segmentation:} Each CSI sample \( \boldsymbol{u_i} \) in the unlabelled batch \( U \) has dimensions \( N_a \times N_s \times N_t \), where \( N_a \) denotes the number of transmitter-receiver antenna links, \( N_s \) denotes the number of subcarriers, and \( N_t \) denotes the number of packets or timestamps. Each sample \( \boldsymbol{u_i} \) undergoes a segmentation process, which results in a set of CSI windows \( \{\boldsymbol{w^i_1}, \ldots, \boldsymbol{w^i_L}\} \). Each window \( \boldsymbol{w^i_t} \) has dimensions \( N_a \times N_s \times N_f \), where \( N_f \) represents the number of CSI frames in each window. The total number of windows \( L \) generated from each sample is calculated by \( L = \frac{N_t}{N_f} \). 
    
$\bullet$ \textbf{Stochastic augmentations:} In this phase, each CSI window $\boldsymbol{w_t}$ undergoes random transformations through a customized suite of augmentations designed specifically for CSI time-series data including \textit{Gaussian noise}, \textit{time flip}, \textit{time mask}, \textit{subcarrier mask}, and a novel augmentation, named \textit{dual view}, proposed specifically for wireless sensing. This set of augmentations  $\mathcal{T}$ yields two unique distorted versions of the data sample, denoted as $\boldsymbol{x_t^A}$ and $\boldsymbol{x_t^B}$.
    
$\bullet$ \textbf{Latent representation generation using base encoder:} A base encoder, $E_{\theta}$ is employed to generate a sequence of latent representations from the windows. Each of the two augmented views is processed by two separate encoders with their respective learning parameters $\boldsymbol{\theta^A}$ and $\boldsymbol{\theta^B}$. This yields latent representations $\boldsymbol{z_t^A} = E_{\boldsymbol{\theta^A}}(\boldsymbol{x_t^A})$ and $\boldsymbol{z_t^B} = E_{\boldsymbol{\theta^B}}(\boldsymbol{x_t^B})$, where $\boldsymbol{z_t^A}, \boldsymbol{z_t^B} \in \mathbb{R}^D$ represent the encoder's output in the embedding space of dimension $D$. The main objective of CAPC is to ensure that the representations $\boldsymbol{z}$, produced by the base encoder, are as feature-rich and robust as possible since these representations are used in downstream tasks.
    
$\bullet$ \textbf{Context embedding generation using an autoregressive model:} Following the generation of latent representations, an autoregressive model, $G$, is applied to condense the sequence of latent representations into a singular context embedding for each augmented view. Specifically, $\boldsymbol{c_t^A} = G_{\boldsymbol{\gamma^A}}(\boldsymbol{z_{\leq t}^A})$ and $\boldsymbol{c_t^B} = G_{\boldsymbol{\gamma^B}}(\boldsymbol{z_{\leq t}^B})$ are obtained, where both $\boldsymbol{c_t^A}$ and $\boldsymbol{c_t^B}$ reside in the $\mathbb{R}^H$ space.
Here, the expressions $\boldsymbol{c_t^A}$ and $\boldsymbol{c_t^B}$  represent the process of applying autoregressive model $G$ to the sequence of latent representations $\textbf{$z^A_{\leq t}$}$ and $\textbf{$z^B_{\leq t}$}$ up to a random time index $t$ that is initiated at each epoch for each branch A and B, respectively. This means that the model processes each $\boldsymbol{z_i}$ in the sequence using the hidden state from the previous step and the current latent representation $\boldsymbol{z_i}$, continuing this process until $\boldsymbol{z_t}$. The output at this final step $t$ is the context embedding for each view, denoted as $\boldsymbol{c_t^A}$ and $\boldsymbol{c_t^B}$.
Here, $\mathbb{R}^H$ denotes the embedding space utilized by the model's loss function. The parameters $\boldsymbol{\gamma^A}$ and $\boldsymbol{\gamma^B}$ represent the distinct learning parameters for the autoregressive model's head, corresponding to each view.
    
$\bullet$ \textbf{Hybrid contrastive loss function:} CAPC's core innovation is a hybrid contrastive loss function that combines the CPC ($\mathcal{L}_{CPC}$) \cite{oord2018representation} and Barlow Twins ($\mathcal{L}_{BT}$) \cite{zbontar2021barlow} loss functions. The CPC loss predicts future latent representations from context embeddings $\boldsymbol{c_t^A}$ and $\boldsymbol{c_t^B}$, independently for each branch, enforcing the model to learn the underlying shared information between the windows. Meanwhile, $\mathcal{L}_{BT}$ enhances consistency between $\boldsymbol{c_t^A}$ and $\boldsymbol{c_t^B}$, ensuring robustness against augmentations and preventing dimensional collapse by reducing redundancy within embeddings.

\textbf{Remark:} Note that we use the RSCNet encoder \cite{barahimi2024rscnet} which already includes components like CSI segmentation and an autoregressive model within our framework. However, we substituted the LSTM in the autoregressive block with a gated recurrent unit (GRU) \cite{cho2014learning} due to its greater efficiency compared to LSTM. Alternatively, any standard encoder and autoregressive model, such as Transformers \cite{vaswani2017attention}, could be employed, potentially enhancing performance with extra computational resources.


\subsection{Proposed Model-based Augmentation}

Data augmentation is crucial for the effectiveness of many SSL techniques \cite{grill2020bootstrap}. These methods aim to create distorted views of the same sample and then maximize the agreement between the representations of these views, making the design of appropriate data augmentations essential. Typically, these augmentations are distortions that naturally occur in the input data but do not alter the inherent semantics of the data, namely their label in the downstream task. By applying these augmentations, we force the encoder to stay invariant to these distortions and to learn the fundamental characteristics—specifically, the useful features of the downstream task that are consistent across both views.

We propose a novel augmentation technique for SSL in wireless sensing, termed \textit{Dual View}. This method leverages reciprocal links of wireless transceivers to isolate the free space propagation channel from electronic distortions inherent in the transmission and reception processes. By randomly assigning uplink and downlink CSI data from these reciprocal links to two branches of the network, our technique aims to extract the core characteristics of the over-the-air channel. 

In classical electromagnetics, the reciprocity theorem \cite{vaidyanathan2010signal} asserts that the communication channel between the transmitter and receiver is the same in both directions, regardless of the signal's direction. However, applying this concept straightforwardly may miss a critical subtlety: the channel's reciprocity pertains only to the antenna-to-antenna interaction. Specifically, the theorem addresses the signal's behaviour once it enters the free space medium.
In time division duplex (TDD) systems, such as those used in WiFi networks, the reciprocity theorem holds true as long as the uplink and downlink transmissions occur within the channel's coherence time and use the same frequency band. In contrast, in frequency division duplex (FDD) systems, this reciprocity does not apply because the uplink and downlink transmissions are typically separated by a frequency difference greater than the coherence bandwidth \cite{rottenberg2019channel}. Although full reciprocity does not exist in FDD systems, the uplink and downlink channels do exhibit partial reciprocity properties, such as multipath angle and delay \cite{zhong2020fdd}, which may still provide useful sensing information.

As previously mentioned, WiFi devices use TDD technology, meaning full reciprocity applies to them.
Nevertheless, the CSI measured by standard WiFi devices does not solely reflect the free space channel. Instead, the measured CSI at either end, is the measurement of the channel, all the way from the digital representation of subcarriers in the transmitter, to their digital representation at the receiver. This includes the entire analog medium traversed by the signal, including base-band (low frequency) amplifiers, oscillators, mixers, (HF) power amplifiers, front-end filters, duplexers, matching networks etc. Thus, the signal path includes significant electronic processing inside the transmitter, and similarly at the receiver, where the signal again passes through a front-end filter, low noise amplifier (LNA), mixer, base-band channelizer filter, and baseband amplifiers. These electronics involved in the transmission/reception chain have a non-trivial effect on the signal. These electronics significantly alter the signal, influencing the CSI calculation, which combines the effects of transmitter electronics, free space propagation, and receiver electronics. However, to study phenomena such as sensing or HAR, which only affects the free space propagation channel, it is essential to isolate this segment from the influences of transmitter and receiver effects.

An answer arises by generating a latent representation of the channel that discounts the electronic effects by maximizing the information between the uplink and downlink CSI using a SSL method. The rationale is that the impact of free space propagation represents the common information in the uplink and downlink CSI, and their discrepancies are due to electronic distortions. Therefore, by maximizing information between these two branches, the encoders are forced to overlook the artifacts attributable to the electronics processing chain and instead generate representations that are closely tied to the variations in the over-the-air channel, which are crucial for sensing applications. In Section \ref{sec:experiments}, we demonstrate how this augmentation, coupled with our innovative loss function, achieves such representations highlighted by our superior performance in downstream sensing tasks.

\subsection{Hybrid Contrastive Loss Function}


We present a new hybrid contrastive loss function that enhances the encoders and autoregressive models within each branch. This is achieved by employing the CPC loss function for intra-branch temporal prediction and the Barlow Twins loss function to ensure inter-branch contextual consistency.

\subsubsection{Temporal Prediction}

Following the CPC methodology \cite{oord2018representation}, at each iteration, we select a random timestamp $t \leq L - T$, where $T$ is the number of future windows the model aims to predict based on the windows up to $t$. For each branch, given the augmented input sequence $\boldsymbol{x}$, we generate the latent representation $\boldsymbol{z_t} = E_{\theta}(\boldsymbol{x_t})$ with potentially reduced dimensions. An autoregressive head $G_{\gamma}$ then summarizes all $\boldsymbol{z_{\leq t}}$ into a context latent representation $\boldsymbol{c_t} = G_{\gamma}(\boldsymbol{z_{\leq t}})$.

This context embedding is used to predict each future window $t + k$, where $k \leq T$. However, instead of making a direct prediction, we introduce a log-bilinear model $f$ as a density ratio estimator to preserve the mutual information between $\boldsymbol{x_{t+k}}$ and $\boldsymbol{c_t}$ as:
\begin{equation}
    f_k (\boldsymbol{x_{t+k}}, \boldsymbol{c_t}) = \text{exp} ( \boldsymbol{z^T_{t+k}} \boldsymbol{W_k} \boldsymbol{c_t}  )\text{,}
\end{equation}
where $\boldsymbol{W_k}$ is a linear layer predicting the future window $t + k$ via the transformation $\boldsymbol{W_k} \boldsymbol{c_t}$, with each $\boldsymbol{W_k}$ tailored for every step $k$. This linear transformation ensures the mapping aligns the dimensions of $\boldsymbol{c_t}$ with those of $\boldsymbol{z_{t+k}}$.

The rationale for preferring implicit prediction over explicit approaches, such as mean squared error, is that explicit methods are often computationally intensive. They attempt to model complex relationships within windows of $\boldsymbol{x}$ rather than extracting shared information between $\boldsymbol{x}$ and $\boldsymbol{c}$. Conversely, the CPC method chooses to model the mutual information between the context embedding, which represents current information, and future latent representations through a non-linear mapping, denoted as $f(\boldsymbol{x}, \boldsymbol{c})$. By maximizing the mutual information between the representations, we extract shared features inherent to both the current and future windows.  

To maximize the mutual information and optimize the estimation $f$, we employ the InfoNCE contrastive loss function, which uses categorical cross-entropy to distinguish the positive sample—the correct prediction—from others within the context's latent distribution. For a given batch size $N$, let $X = \{ \boldsymbol{x^1_{t + k}}, \ldots, \boldsymbol{x^N_{t + k}} \}$ represent the set of potential predictions at step $k$ for each batch sample. The CPC loss is thus defined as:
\begin{equation}
    \mathcal {L}_{CPC}= - \frac{1}{T} \sum_{k=1}^{T} \mathbb{E}_X \log \frac{f_k (\boldsymbol{x^i_{t+k}}, \boldsymbol{c^i_{t}})}{\sum_{\boldsymbol{x^j} \in X} f_k (\boldsymbol{x^j}, \boldsymbol{c^i_{t}})}\text{,}
\end{equation}
The use of randomly selected current timestamp, denoted as $t$, for each batch aims to enhance the generalization capabilities of the representations. CPC is applied independently within each branch of the twin network but with the same current timestamp for both branches as well as the same transformation $\boldsymbol{W}$, thereby predicting future latent representations $\boldsymbol{z^A_{t+k}}$ and $\boldsymbol{z^B_{t+k}}$ based on their respective context latent representations $\boldsymbol{c^A_t}$ and $\boldsymbol{c^B_t}$, respectively, using the $\mathcal{L}_{CPC}$ loss.

\subsubsection{Contextual Consistency}
To further enhance representation robustness and reduce redundancy, the Barlow Twins loss function \cite{zbontar2021barlow} is utilized to preserve the uniqueness of features while maintaining their invariant nature across different augmentations. The loss function is employed on the context latent representations $\boldsymbol{c^A_t}$ and $\boldsymbol{c^B_t}$ and is formulated as follows:

\begin{equation}
    \mathcal{L}_{BT} = \underbrace{\sum_{i=1}^{H} (\boldsymbol{\mathcal{C}}_{ii} - 1)^2}_\text{{invariance term}} + \underbrace{\lambda \sum_{i,j=1}^{H}\boldsymbol{\mathcal{C}}_{ij}^2\text{,}}_{\text{redundancy reduction term}}
\end{equation}
Here, $\lambda$ is a positive constant that balances the importance of the two terms of the loss function, and 
$\boldsymbol{\mathcal{C}}$ is the cross-correlation matrix computed between the outputs $\boldsymbol{c^A}$ and $\boldsymbol{c^B}$ of the two identical networks along the batch dimension. The cross-correlation matrix $\boldsymbol{\mathcal{C}}$ is defined as:
\begin{equation}
    \boldsymbol{\mathcal{C}}_{ij} = \frac{\sum_{b} \boldsymbol{c}_{i}^{b,A} \boldsymbol{c}_{j}^{b,B}}{\sqrt{\sum_{b} (\boldsymbol{c}_{i}^{b,A})^2 \sum_{b} (\boldsymbol{c}_{j}^{b,B})^2}}\text{,}
\end{equation}
In this equation, \( b \) indexes batch samples, and \( i, j \) indexes the feature embeddings of the context latent representations. The values of \( \boldsymbol{\mathcal{C}} \) range from $-1$ (perfect anti-correlation) to 1 (perfect correlation). The Barlow twins' invariance term aims to make the embedding invariant to any applied augmentations by setting the diagonal elements of the cross-correlation matrix to 1. Meanwhile, the redundancy reduction term seeks to decorrelate the various vector components of the embedding by setting the off-diagonal elements of the cross-correlation matrix to zero. This process of decorrelation minimizes redundancy among the output units, ensuring that they hold distinct information about the sample.

Contrary to the original Barlow Twins method, which incorporates a projection head to map the latent representations to a separate embedding space for loss function optimization, our approach applies the Barlow Twins loss directly to the context representations. In other words, the autoregressive model in our framework also serves as the projection map.

\subsubsection{Hybrid Loss} Our hybrid loss function integrates these components, leveraging a hyperparameter $\beta$ to align the scale of the CPC loss, $\mathcal{L}^A_{CPC}$ and $\mathcal{L}^B_{CPC}$, to the Barlow Twins loss $\mathcal{L}_{BT}$, resulting in a balanced and effective optimization criterion for our SSL framework:
\begin{equation}
\mathcal{L} = \mathcal{L}_{BT} + \beta (\mathcal{L}^A_{CPC} + \mathcal{L}^B_{CPC})\text{,}
\end{equation}

Most SSL methods, including CAPC, extract valuable representations by minimizing the distances between embedding vectors of augmented samples. However, without additional mechanisms, these methods could lead to a \textit{complete collapse} solution, where the resultant representations become constant across different inputs. Furthermore, they may also experience \textit{dimensional collapse}, where the embedding vectors are confined to a lower-dimensional subspace, failing to utilize the entire embedding space.
 In CAPC, several components aid in preventing both complete and dimensional collapse. The use of negative samples in the CPC loss function prevents the complete collapse by ensuring that negative samples are distanced from positive samples, thus introducing variance in the representations.  Moreover, the redundancy reduction term of Barlow Twins not only prevents complete collapse, but also combats dimensional collapse by promoting the decorrelation of feature embeddings. This hybrid strategy effectively enhances the robustness and utility of the learned representations.

\begin{algorithm}[h]
\caption{Pseudocode for CAPC}
\SetAlgoLined
\KwIn{Unlabelled CSI batch $U = \{ \boldsymbol{u_i} \}_1^N$}
\KwOut{Updated model parameters $\boldsymbol{\theta^A}, \boldsymbol{\theta^B}, \boldsymbol{\gamma^A}, \boldsymbol{\gamma^B}, \boldsymbol{W_{1 \ldots k}}$}
\While{$\text{step} < \text{total steps}$}{
    Segment each input $\boldsymbol{u}$ to create CSI windows $\{ \boldsymbol{w_t} \}_1^L$\;
    Apply augmentation $\mathcal{T}$ to each window $\boldsymbol{w}$ generating $\boldsymbol{x^A}, \boldsymbol{x^B}$\;
    Select a random timestamp $t \leq L - T$\;
    Extract latent representation $\boldsymbol{z^A} = E_{\boldsymbol{\theta^A}}(\boldsymbol{x^A})$ and $\boldsymbol{z^B} = E_{\boldsymbol{\theta^B}}(\boldsymbol{x^B})$ for $\boldsymbol{x_1}, \dots, \boldsymbol{x_{t+T}}$\;
    Summarize the representations of the first $t$ windows, $\boldsymbol{c^A_t} = G_{\boldsymbol{\gamma^A}}(\boldsymbol{z^A_{\leq t}}) \text{ and } \boldsymbol{c^B_t} = G_{\boldsymbol{\gamma^B}}(\boldsymbol{z^B_{\leq t}})$\;
    Calculate the mutual information between the context prediction $\boldsymbol{c^A_t}$ or $\boldsymbol{c^B_t}$, and each possible prediction $\boldsymbol{x}$ by  $f_k (\boldsymbol{x}, \boldsymbol{c_t}) = \text{exp} ( E_{\theta}(\boldsymbol{x}) \boldsymbol{W_k} \boldsymbol{c_t}  )$\;
    Update $\boldsymbol{\theta^A}, \boldsymbol{\theta^B}, \boldsymbol{\gamma^A}, \boldsymbol{\gamma^B}, \boldsymbol{W_{1 \ldots k}}$ by minimizing $\mathcal{L}_{BT} + \beta (\mathcal{L}^A_{CPC} + \mathcal{L}^B_{CPC})$\;
}
\end{algorithm}

\section{Experimental Settings and Evaluation}
\label{sec:settings}

In this section, we present the dataset specifications, baselines for comparison, evaluation criteria, training configuration of the neural network architecture, and the augmentations used in addition to the proposed dual view. 

\subsection{Datasets}

\subsubsection{SignFi}

We employed the SignFi gesture recognition dataset \cite{10.1145/3191755}, specifically designed for sign language gesture recognition tasks. This dataset features a significant volume of data instances but has a limited number of samples per class due to its extensive variety of sign language words (classes), totalling 276. The SignFi dataset, acquired through the Intel 5300 NIC \cite{10.1145/1925861.1925870} in a single-transmitter single-receiver setup, consists of 3 antennas at the receiver, a single antenna for the transmitter, 30 subcarriers, and 200 packets per data instance, resulting in each sample possessing $3 \times 30 \times 200$ dimensions. Its key attributes include:\\
\textbf{(1)} \textbf{Multiple environment setups:} SignFi dataset comes from two different environments: a home and a lab. This variety helps us test how well our method adapts to new environments. We used the lab dataset as the unlabelled dataset for the SSL pre-training as it contained more samples and the home dataset with labels for the supervised evaluation.  \\
\textbf{(2)}  \textbf{Dual view CSI:} SignFi dataset includes synchronized uplink and downlink CSI for each sample in the dataset allowing us to utilize them for dual view augmentation in SSL pre-training. To the best of our knowledge, SignFi is the only database offering synchronized uplink and downlink CSI. \\
\textbf{(3)}  \textbf{Substantial Sample Volume:} SignFi dataset comprises 5520 instances from the lab environment (20 samples per class) and 2760 instances from the home environment (10 samples per class), making it a notably large dataset. \\
\textbf{(4)}  \textbf{User Diversity:} Lab environment samples are collected independently for each of the 5 users participating in the experiments, meaning that at any given time, only one person is performing an activity in the room. In contrast, home environment samples are collected from a single user.

\subsubsection{UT HAR}

We also employed the UT HAR dataset \cite{8067693} for evaluating transfer learning. Specifically, we tested the RSCNet backbone encoders, which were pre-trained using unlabelled CSI data in the SignFi lab, on a subset of labelled samples from UT HAR. These experiments aimed to assess the effectiveness of the CAPC representations for adapting to new tasks and environments.

The UT HAR data was collected using the Intel 5300 NIC \cite{10.1145/1925861.1925870}, similar to the setup used in SignFi.  The data samples have dimensions of 3 antenna links, 30 subcarriers, and 250 packets. Given that the antenna and subcarrier dimensions align with those of SignFi, and considering our use of a windowing size $N_f = 10$ in the encoders for both CAPC and all baseline models, we were able to use the same pre-trained encoders from SignFi without requiring additional preprocessing.

Additionally, this dataset, which consists of samples all collected from one user, categorizes human activities into seven types: lying down, falling, walking, running, sitting down, standing up, and empty room.  The choice to use this dataset solely for fine-tuning and transfer learning purposes is due to its relatively small size, with only 3,977 samples in the training set.


\subsection{Evaluation Criteria}

Our evaluation criteria shown in Figure~\ref{fig:supervised} are designed to assess how effectively a pre-trained encoder with SSL adapts to downstream WiFi sensing tasks. Specifically, we aim to understand the applicability of the encoder's representations when confronted with a limited number of labelled samples from the downstream task. We base our investigation on two main questions:

\begin{figure}[h]
    \centering
    \includegraphics[width=1\linewidth]{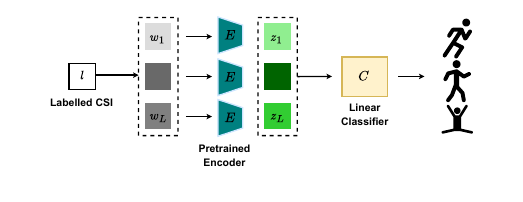}
    \caption{\textbf{Supervised evaluation:} A linear classifier $C_{\boldsymbol{\phi}}$ is fine-tuned with labelled CSI based on the concatenated representations from all windows generated by the pre-trained encoder $E_{\boldsymbol{\theta^A}}$. The pre-trained encoder's weights $\boldsymbol{\theta^A}$ are frozen in linear classification but fine-tuned in the semi-supervised evaluation.}
    \label{fig:supervised}
\end{figure}

\textbf{(1)} \textbf{Can the SSL encoder extract relevant features for the downstream task without additional adaptation?} We analyze if the representations produced by the encoder are sufficient and generalizable for the downstream tasks without additional data-specific training. The encoder's weights, denoted as $\boldsymbol{\theta^A}$, are frozen to prevent the infusion of task-specific information. The latent representations, generated by the encoder for each data window, are concatenated and used as input for a linear classifier $C_{\boldsymbol{\phi}}$. This classifier is trained using the cross-entropy loss function, defined by:
\begin{equation}
\mathcal{L}_{c}(\boldsymbol{x}, \boldsymbol{y}) = -\mathbb{E}_{(\boldsymbol{x}, y)} \sum_{i=1}^{n} \left [ \mathbb{I}[y = i] \log \left( C_{\boldsymbol{\phi}} \left( E_{\theta^A}(\boldsymbol{w_{1, \cdot\cdot\cdot, L}}) \right) \right) \right]\text{,}
\label{eq:cross_entropy}
\end{equation}
Here, $n$ is the number of class labels, and $E_{\boldsymbol{\theta^A}}(\boldsymbol{w_{1, \cdot\cdot\cdot, L}})$ represents the merged embeddings generated by the encoder across all CSI windows of each sample i.e. $\{ \boldsymbol{w_t} \}_1^L$. 
This setup tests the hypothesis of how a well-trained encoder can enable a linear classifier to perform effectively on downstream tasks without the additional information of the task in the representations.

\textbf{(2)} \textbf{Does partial fine-tuning of the encoder enhance its adaptability to the downstream tasks?} We examine whether adjusting the weights of the encoder, $\boldsymbol{\theta^A}$, enhances its performance and adaptability to the specific characteristics of the downstream task. This creates a balance between leveraging the learned representations in the self-supervised mode and utilizing available labelled samples.

Unlike in linear evaluation where the encoder is completely frozen, in this semi-supervised evaluation setup, both the encoder and the classifier are trained, but the encoder is subjected to a lower learning rate compared to the classifier as well as fewer training epochs in general. This method allows the encoder to fine-tune its pre-learned representations to the specifics of the new task without significantly drifting from its original, generalizable features. Importantly, this approach helps prevent catastrophic interference, where the model could otherwise forget previously learned information upon acquiring new, potentially biased information from the limited number of labelled samples in the downstream task. This strategy ensures a robust evaluation of the SSL method's effectiveness and adaptability. 

\subsection{Baselines}

To thoroughly evaluate the effectiveness of our proposed method, CAPC, we conducted comparisons with four established SSL methods. Specifically, we compared CAPC against: \textit{(1) SimCLR} \cite{chen2020simple}, which serves as a standard contrastive SSL benchmark; \textit{(2) Barlow Twins} \cite{zbontar2021barlow} and \textit{(3) CPC} \cite{oord2018representation}, both included as ablation studies due to their integration within CAPC's framework; \textit{(4) AutoFi} \cite{yang2022autofi}, a recent SSL method tailored for CSI WiFi sensing. Furthermore, we compare CAPC against a fully supervised training model where the encoder $E_{\boldsymbol{\theta}}$ is randomly initialized and trained with the labelled data from scratch. This comparison is intended to highlight the performance gains of SSL even with limited training labels.

\textbf{Remark:} To ensure a fair comparison, all methods utilized the same backbone encoder architecture, namely the RSCNet encoder \cite{barahimi2024rscnet}. Additionally, unlike CAPC and CPC, which employ an autoregressive model during pre-training, SimCLR, Barlow Twins, and AutoFi use a projector to independently map each window’s representation into an embedding space for applying their respective loss functions. Specifically, we adopted the Barlow Twins' projector configuration, which consists of three fully connected layers paired with ReLU activation functions.
By using the same backbone encoder across all methods, we maintained consistent model complexity, as detailed by the FLOP counts in Figure~\ref{fig:window_size}. This consistency ensures that any differences in representation quality are solely attributable to SSL methodology and augmentations rather than model complexity, thus guaranteeing a fair and balanced comparison among CAPC and the baseline methods.


\begin{table*}[h]
\caption{\textbf{Evaluation of representations fine-tuned on SignFi Home} derived from pre-trained CAPC and baseline SSL methods on: (1) linear classification using frozen representations; (2) semi-supervised classification with fine-tuned representations. pre-training occurs in the lab environment of SignFi, and supervised evaluations are conducted in various fractions of the home environment. We report the accuracies (in \%) of supervised evaluation, highlighting the best results in \textbf{bold} and the second-best results with \underline{underlining}. All methods use RSCNet encoder thus having the same number of parameters and complexities.}
\centering
\begin{threeparttable}
    \centering
    \begin{tabular}{ll|cccccc|c}
        \toprule
        \multicolumn{1}{c}{} & \multicolumn{6}{c}{Shots} & \multicolumn{1}{c}{} \\
        \cmidrule(r){3-8}
        Evaluation Method & Method & 2 & 4 & 6 & 8 & 10 & 12 & Avg. \\
        \midrule
        \multirow{7}{*}{Linear}
        & Supervised & - & 57.97 & 78.99 & 89.95 & 82.79 & 91.12 & 66.80 \\  
        & SimCLR & 59.6 & 82.25 & 92.57 & 95.02 & 96.01 & \textbf{98.01} & 87.24 \\
        & CPC & 55.89 & 75.82 & 86.23 & 89.49 & 93.57 & 94.66 &  82.61 \\
        & Barlow Twins & 54.08 & 76.00 & 92.66 & 92.39 & 95.29 & 96.92 & 84.56 \\
        & AutoFi & 59.15 & 79.62 & \underline{92.84} & \textbf{95.92} & \underline{96.65} & 97.55 & 86.95 \\
        & \textbf{CAPC\tnote{*}} & \underline{63.41} & \underline{85.51} & 92.48 & 95.38 & 96.56 & \underline{97.83} & \underline{88.53} \\
        & \textbf{CAPC} & \textbf{65.67} & \textbf{88.50} & \textbf{93.84} & \underline{95.83} & \textbf{97.55} & 97.55 & \textbf{89.82} \\
        \midrule
        \multirow{7}{*}{Semi-Supervised}
        & SimCLR & \underline{51.27} & \underline{75.18} & \underline{89.13} & \underline{93.57} & 94.84 & 96.38 & \underline{83.39} \\
        & CPC & 46.01 & 67.03 & 78.62 & 87.59 & 91.03 & 94.47 & 77.46 \\
        & Barlow Twins & 46.74 & 68.93 & 78.71 & 89.40 & 94.02 & 95.29 & 78.85 \\
        & AutoFi & 46.56 & 72.19 & 86.50 & 92.93 & \underline{96.11} & 96.47 & 81.79 \\ 
        & \textbf{CAPC*} & 49.73 & 72.83 & 87.5 & 93.12 & 95.83 & \underline{97.74} & 82.79 \\
        & \textbf{CAPC} & \textbf{57.52} & \textbf{82.52} & \textbf{92.57} & \textbf{96.47} & \textbf{97.19} & \textbf{97.92} & \textbf{87.36} \\
        \bottomrule
    \end{tabular}%
    \begin{tablenotes}
        \item[*] Refers to the CAPC model being trained with the second best combination of augmentations, noise and subcarrier mask, instead of noise and dual view.
    \end{tablenotes}
     \end{threeparttable}
    \label{tab:results}
\end{table*}

\subsection{Augmentations}

In addition to the proposed dual view augmentation, we evaluated several common augmentations used in the context of time-series \cite{pmlr-v136-mohsenvand20a, 9930826} and WiFi sensing \cite{9522151}: \\
$\bullet$ \textbf{Gaussian Noise:} introduces random Gaussian noise with zero mean and 0.1 standard deviation into data, simulating inherent noise in CSI, and enhancing model resilience.\\
$\bullet$ \textbf{Time Flip:} flips CSI samples along the time axis to accommodate palindromic activities, where time-reversed patterns represent the same activity.\\
$\bullet$ \textbf{Time Mask:} randomly masks a segment of the time dimension in each CSI window, varying its location. This teaches the model to predict missing temporal information, simulating scenarios with temporal disruptions or signal losses.\\
$\bullet$ \textbf{Subcarrier Mask:} masks a set of subcarriers in each CSI sample, forcing the model to infer activities using the remaining subcarriers. This augmentation improves the handling of frequency-selective fading or interference and underscores different subcarriers' significance in activity recognition.

{\bf Choosing best augmentations for each method:} We conduct a comprehensive analysis of the aforementioned augmentations, both individually and in combination, on the performance of our method and the baselines. This allowed us to identify the optimal augmentation mix for each method and to demonstrate the robustness of each method in response to the various augmentations. The augmentations chosen for each method are in Figure~\ref{fig:augmentations}. For CAPC, the chosen augmentations are dual view and noise. For SimCLR, it is time mask. The Barlow Twins use noise and time mask, while AutoFi employs noise and time flip. CPC relies solely on prediction for its SSL task, thus not using any augmentations.

\textbf{Remarks:} In experiments where the dual view augmentation was not applied, we treated the uplink and downlink CSI as separate samples, effectively doubling the dataset size during the SSL phase. This approach highlights the flexibility of our framework, showing that it is not limited to using dual view augmentation and can still operate effectively with either uplink or downlink CSI samples. Unlike dual view augmentation, which requires both CSI samples, CAPC can function effectively without needing both.

\subsection{Training Configuration}

Our model, developed using PyTorch, employs the LARS optimizer during the SSL phase \cite{DBLP:journals/corr/abs-1708-03888}. The training lasts for 300 epochs with a batch size of 128. We initialize the learning rate for weights at 0.2 and for biases and batch normalization parameters at 0.0048. The initial 10 epochs act as a warm-up period for the learning rate, which is then reduced following a cosine decay schedule \cite{DBLP:journals/corr/LoshchilovH16a}. The weight decay is configured at $1.5 \times 10^{-6}$. The trade-off parameter of Barlow Twins loss is set to $\lambda = 0.002$ and the trade-off parameter of CAPC $\beta$ is set to 50 to scale the loss terms. These configurations largely follow those reported in \cite{zbontar2021barlow}. In our model, unlike \cite{zbontar2021barlow}, the weights between the twin networks are not shared to enhance performance. Regarding the number of future windows prediction $T$ in CAPC and CPC, we chose $9$ for CAPC and $2$ for CPC, as these values show better performance for each method as shown in Figure~\ref{fig:window_size}.

During the evaluation phase, we switch to the Adam optimizer \cite{kingma2014adam}, maintaining a batch size of 512. For linear evaluation, the learning rate starts at $10^{-2}$ and follows a cosine decay schedule over 100 epochs, training only the linear classifier. In the semi-supervised evaluation, the configuration remains the same except the model is trained for only 20 epochs, and the encoder's learning rate is set at $5 \times 10^{-3}$, following a similar decay schedule. 

For the model architecture, we use a window size $N_f = 10$, an encoder embedding size $D = 128$, and 128 nodes in the GRU autoregressive model's hidden layer, the projection hidden layers and the embedding size $h$. The number of nodes in the hidden layer of the linear classifier $C_{\phi}$ is half of its input size, which equals the embedding size $D$ multiplied by the sequence length $L$.

\section{Results and Discussions}
\label{sec:experiments}

\subsection{Supervised evaluation of representations under different number of labelled samples}
\label{subsec:database-size-evaluation}

We conduct extensive experiments to evaluate the effectiveness of our proposed CAPC model across different portions of the labelled SignFi Home dataset. The evaluation involved fine-tuning the pre-trained encoder using 2 to 12 samples per class, or \textit{shots}. Table~\ref{tab:results} presents the results of our fine-tuned CAPC model compared to baseline methods, for both linear and semi-supervised evaluations. Additionally, we include the results of fully supervised training for reference. Other than CAPC with noise and dual view augmentation, which proved to be the most effective combination, we also tested CAPC with noise and subcarrier mask augmentations. These experiments showcase the superiority of our proposed dual view augmentation and demonstrate that our model can also deliver strong performance with alternative transformations.

In terms of linear evaluation, our method consistently outperforms all other compared methods, achieving the highest average accuracy of 89.82\%. In few-shot scenarios (2 and 4 shots), both versions of CAPC achieved the highest accuracies, with the dual view version reaching 65.67\% and 88.50\% for 2 and 4 shots respectively. The best baseline, SimCLR, achieved 59.6\% and 82.25\% for the same scenarios. For higher numbers of shots, the performance differences among all SSL methods, except CPC, become less pronounced, yet CAPC still records the highest accuracies in most scenarios, falling short by less than 0.5\% only in the 8 and 12 shots categories. Generally, CAPC with either augmentation combination outperforms other methods by approximately 2.6\% in average accuracy. Furthermore, we observe that supervised learning alone performs poorly across all shots compared to SSL methods, particularly in few-shot scenarios. For instance, with just 2 shots, the supervised model fails to converge and achieve meaningful performance. Furthermore, the average accuracy of supervised training is approximately 23\% lower than that of CAPC.

In semi-supervised evaluations, CAPC with dual view augmentation distinctly surpasses all other SSL methods by a substantial margin of 4\%. This augmentation also appears to enhance semi-supervised training significantly, indicating that the encoder weights from the CAPC with dual view are particularly well-suited for fine-tuning. In few-shot scenarios, the performance gap between our method and the top baseline, SimCLR, is even more pronounced, with about a 7\% difference in the 2 and 4 shot scenarios. CAPC without dual view augmentation also exceeds all baselines except for SimCLR, which shows comparable performance with only a 0.6\% gap.

Overall, the results underscore the critical role of leveraging both temporal dependencies and wireless channel characteristics for SSL, especially in few-shot scenarios.

\subsection{Transfer learning for a different task}

Transfer learning involves adapting a pre-trained model to new tasks across different domains, demonstrating the model's ability to generalize well beyond its original training context. We applied the same pre-trained RSCNet backbone encoders—originally trained on the unlabelled SignFi lab dataset for sign language gesture recognition—to the UT HAR dataset. For this new HAR task, we kept the encoders fixed and fine-tuned a linear classifier using only 10 and 20 shots subsets of UT HAR dataset.

Our results, presented in Table~\ref{tab:transfer-learning}, highlight the superior performance of the CAPC approach over other SSL methods and traditional supervised learning for the HAR task. Notably, CAPC achieves comparable results to other baselines even without dual view augmentation. However, integrating dual view augmentation significantly boosts the model’s generalizability and performance, surpassing the next best SSL method, SimCLR, by 1.8\%, and achieving a 54.2\% accuracy in the 10 shots scenario, where traditional supervised learning did not converge to a comparable performance level. These findings highlight the effectiveness of CAPC in cross-task and cross-environment WiFi sensing applications.

\begin{table}[h]
\caption{\textbf{Transfer learning on UT HAR.} Evaluation of adapting SSL pre-trained RSCNet backbone encoders using CAPC and other SSL methods from SignFi's sign language detection task to the HAR task by linear evaluation on top of fixed representations with limited labelled data—10 and 20 shots, constituting about 1.8\% and 3.5\% of available samples, respectively.}
\centering
\begin{tabular}{l|cc|c}
    \toprule
    \multicolumn{1}{c}{} & \multicolumn{2}{c}{Shots} & \multicolumn{1}{c}{} \\
    \cmidrule(r){2-3}
    Method & 10 & 20 & Avg. \\
    \midrule
    Supervised & 8.0 & 56.0 & 32.0 \\
    AutoFi & 51.2 & 55.0 & 53.1 \\
    CPC & 52.2 & 54.4 & 53.3 \\
    Barlow Twins & \underline{52.8} & 56.0 & 54.4 \\
    SimCLR & 52.6 & 57.2 & \underline{54.9} \\
    \textbf{CAPC*} & 49.8 & \underline{57.4} & 53.6 \\
    \textbf{CAPC} & \textbf{54.2} & \textbf{59.2} & \textbf{56.7} \\
    \bottomrule
\end{tabular}
\label{tab:transfer-learning}
\end{table}

\subsection{Augmentations selection}
\label{subsec:augmentaions}


\begin{table}[h]
\centering
\caption{Summary of Figure~\ref{fig:augmentations}: Presents the average accuracies achieved using various augmentations across different methods.}
\label{tab:augmentaions}
\begin{tabularx}{\linewidth}{l|ccccc|c}
\toprule
Method & \thead{\textbf{Dual}\\\textbf{View}} & \thead{Time\\Mask} & \thead{Subcarrier\\Mask} & \thead{Time\\Flip} & \thead{Noise} & \thead{Avg.} \\
\midrule
AutoFi                    & 49.98   & 80.16            & 79.64                   & 80.67             & 82.97 & 74.68         \\
Barlow Twins     & 79.22    & 87.73           & 86.76                   & 84.00             & 87.77 & 88.10          \\
SimCLR           & \underline{86.50}  & \textbf{89.33}           & \underline{88.60}                   & \underline{89.04}              & \underline{88.95} & \underline{88.60}        \\
\textbf{CAPC}              & \textbf{91.16} & \underline{87.41}             & \textbf{89.67}                   & \textbf{89.11}             & \textbf{91.89} & \textbf{89.85}         \\
\midrule
Avg.           & 76.72      & \underline{86.16}        & 86.17                  & 85.71            & \textbf{87.89} & 84.53       \\
\bottomrule
\end{tabularx}
\end{table}

\begin{figure*}[ht]
    \centering
    \includegraphics[width=1\linewidth]{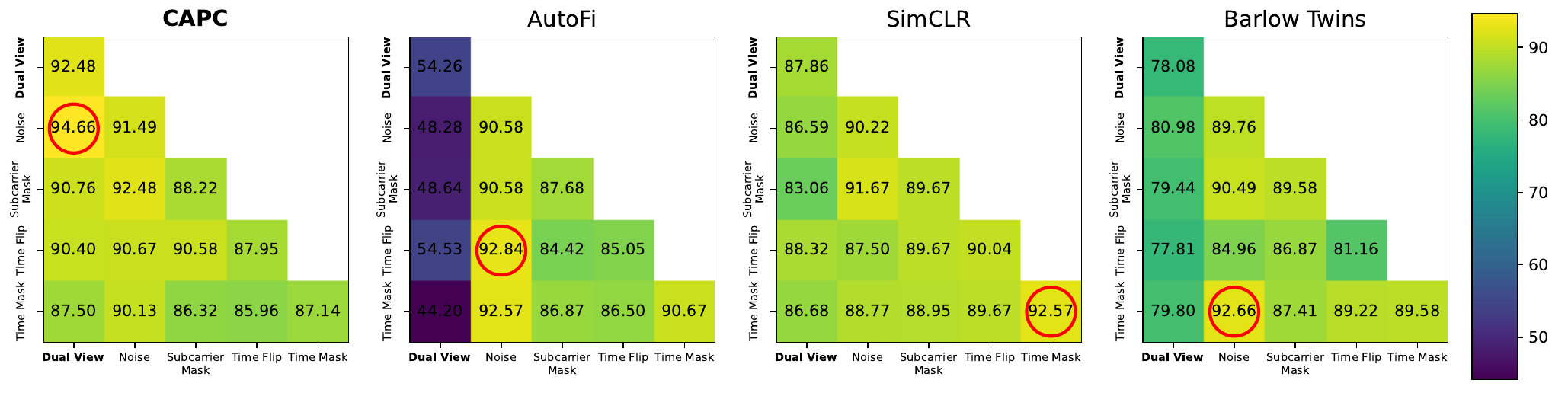}    
    \caption{Linear evaluation of individual and compositional data augmentations. Each diagonal element represents the effect of a single transformation, while off-diagonal elements illustrate the combined impact of two sequentially applied transformations. We report the accuracies (in \%) with 6 shots in the labelled dataset. Red circles indicate the best combination of augmentations.}
    \label{fig:augmentations}
\end{figure*}

The selection of appropriate augmentations is critical in SSL techniques, as these methods are highly sensitive to augmentation choices. While some studies have examined the impact of augmentations in computer vision \cite{chen2020simple}, the exploration of effective augmentations for SSL that are well-suited to CSI data and sensing applications remains limited. Our study focuses on identifying suitable augmentations for both our CAPC method and established baselines. We explored five types of augmentations: dual view, noise, time flip, time mask, and subcarrier mask. These were analyzed both in combination and individually, as demonstrated in Figure~\ref{fig:augmentations}, and the average performance of each augmentation across all methods was summarized in Table~\ref{tab:augmentaions}.

Our comprehensive analysis reveals that noise augmentation consistently enhances model generalization by accommodating inherent data disturbances, making it the most effective augmentation on average. It was part of the best augmentation set for each method, except for SimCLR, which performed better with the time mask alone. In general, our CAPC method outperformed other approaches across all augmentations, except for time mask, where SimCLR showed superior performance. This is likely because the temporal prediction component of CAPC complicates the task when a time mask is also applied, hindering the model’s ability to predict future windows.

Notably, CAPC excelled with the dual view augmentation, achieving an average accuracy of 91.16\%, compared to 86.50\% for SimCLR. This observation is significant as dual view augmentation generally underperformed across other baseline methods. CAPC's remarkable success with this augmentation indicates its unique capability to mitigate electronic distortions, a feature that was anticipated but not as effectively harnessed by other methods such as AutoFi. The distinct advantage of CAPC with dual view augmentation underscores its potential in leveraging complex augmentations that other models fail to utilize effectively.

\subsection{Comparative analysis of contextual loss}
\label{subsec:contextual-loss}

\begin{figure}[h]
    \centering
    \includegraphics[width=0.85\linewidth]{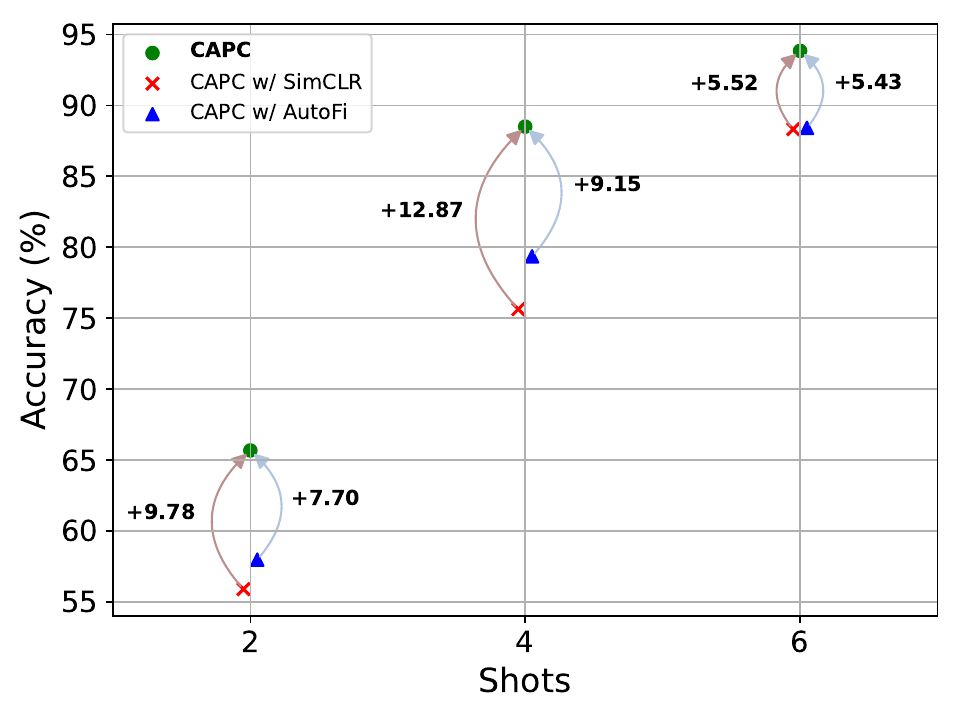}
    \caption{A comparative study of the proposed CAPC method. The CAPC w/ SimCLR and AutoFi mean that we have replaced the Barlow Twins loss in our design with SimCLR and AutoFi, respectively. Showcasing that Barlow Twins has superior performance for enforcing context embedding consistency. We report the experiments under linear evaluation of SignFi Home dataset with 2, 4, and 6 shots.}
    \label{fig:context}
\end{figure}

One of the key novelties of our CAPC method is the integration of hybrid temporal and contextual losses. The use of prediction and autoregressive models to capture temporal dependencies is well-recognized. However, the rationale behind choosing a specific type of contextual loss might not be immediately apparent as various contrastive \cite{chen2020simple} and non-contrastive methods \cite{zbontar2021barlow}, \cite{bardes2022vicreg}, \cite{yang2022autofi} are available. Our investigation shown in Figure~\ref{fig:context} included three distinct loss functions: Barlow Twins, as utilized in our framework, along with SimCLR, and AutoFi. Our findings indicate that the non-contrastive loss from Barlow Twins consistently outperforms both the contrastive loss used in CAPC with SimCLR and the loss of AutoFi across various training scenarios, often by a significant margin. For example, in a scenario with four shots, CAPC equipped with Barlow Twins achieved an accuracy of 88.50\%, markedly higher than the 79.35\% and 75.63\% seen with AutoFi and SimCLR, respectively. Additionally, AutoFi consistently outperforms SimCLR by an average of 2\%. These results suggest that non-contrastive loss functions serve as more effective contextual losses for our proposed CAPC method.

\begin{figure*}[h]
    \centering
    \begin{subfigure}{.32\textwidth}
        \centering
        \includegraphics[width=\linewidth]{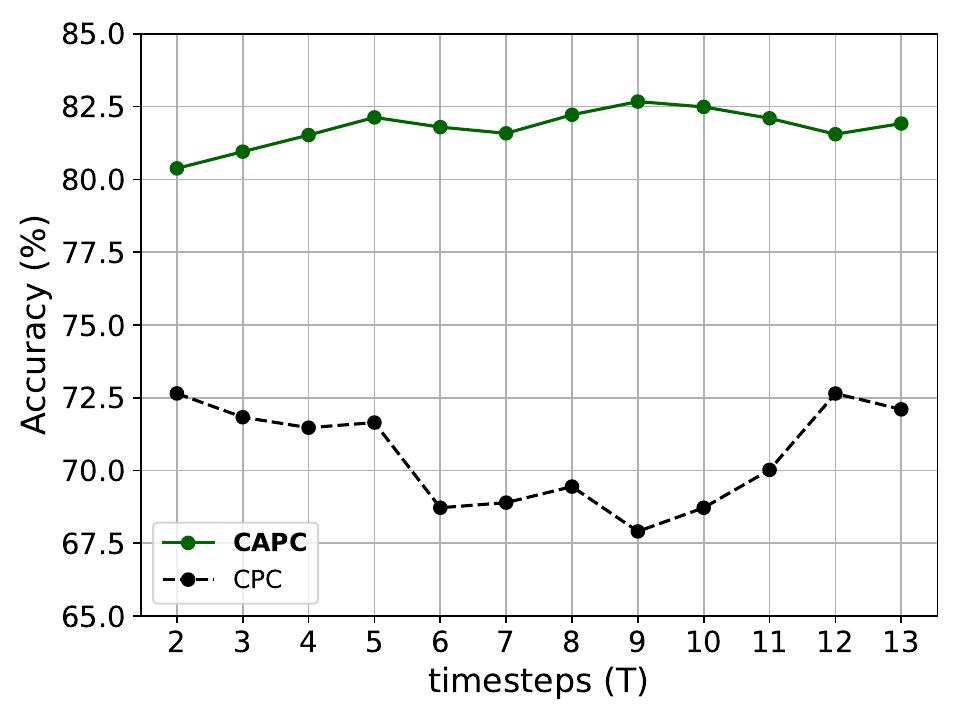}
        \caption{ }
        \label{fig:T}
    \end{subfigure}\hfill
    \begin{subfigure}{.32\textwidth}
        \centering
        \includegraphics[width=\linewidth]{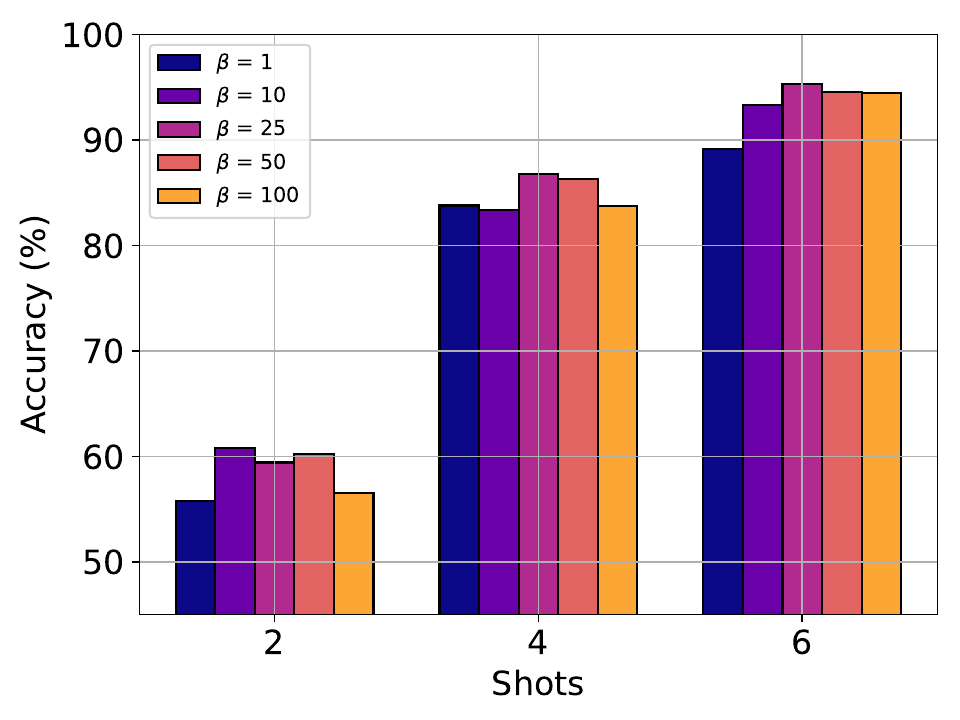}
        \caption{ }
        \label{fig:beta}
    \end{subfigure}\hfill
    \begin{subfigure}{.32\textwidth}
        \centering
        \includegraphics[width=\linewidth]{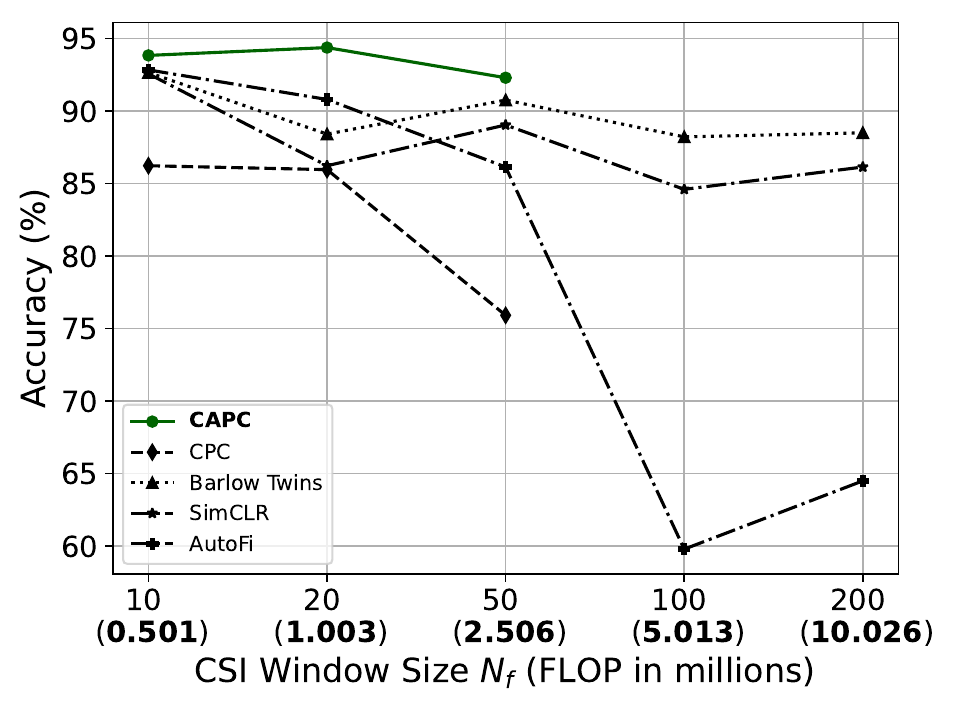}
        \caption{ }
        \label{fig:window_size}
    \end{subfigure}
    \caption{Sensitivity analysis experiments on SignFi dataset: (a) Illustrates how the accuracy of linear evaluation is affected by varying the number of predicted future windows ($T$) for 2, 4, and 6 samples per class, highlighting that CAPC is significantly more stable across different values of $T$ compared to CPC; (b) Depicts the influence of the coefficient $\beta$ on CAPC's performance under linear evaluation; (c) Examines the impact of different window sizes, $N_f$, on the accuracy of CAPC and baseline methods during linear evaluation for 6 samples per class. It also shows the computational complexity of the encoder with varying window sizes. Here, $T = 2$ for both CAPC and CPC due to constraints imposed by the limited number of windows at higher window sizes ($T \leq L - 2$).}
    \label{fig:sensitivity_analysis}
\end{figure*}

\subsection{Hyperparameter selection and sensitivity analysis}
\label{subsec:sensitivity-analysis}

We conducted sensitivity analyses on three primary hyperparameters within the CAPC framework: (1) we illustrate the impact of the number of future windows or timesteps, $T$, predicted by CAPC during SSL, on linear evaluation; (2) we demonstrate the effect of the trade-off parameter, $\beta$, which represents the weight of $\mathcal{L}_{CPC}$ in the loss function; (3) we depict the influence of the number of frames per CSI window, $N_f$, and encoder complexity comparing CAPC to baseline SSL methods. 

The analysis in Figure~\ref{fig:T} reveals the impact of $T$ on the performance of CAPC and CPC. For both frameworks, we present the linear evaluation performance averaged across 2, 4, and 6 shot scenarios in SignFi Home. Both methods exhibit fluctuations; however, CAPC demonstrates greater robustness and fewer fluctuations across different $T$ values, outperforming CPC by approximately 11.3\% on average. Generally, CAPC achieved better results with higher $T$ values, particularly at $T = 9$. Conversely, CPC's performance initially declined with increasing $T$ values but improved upon reaching $T = 12$ and $T= 13$, with $T = 2$ being the optimal value. This implies that CAPC may be more effective at capturing common information over an extended number of windows.

We further investigated the sensitivity of CAPC to the hyperparameter $\beta$, which balances the significance of temporal and contextual consistency in the embeddings. We found that CAPC is relatively insensitive to variations in $\beta$, with values of 25 and 50 both demonstrating slightly improved performance overall, as illustrated in Figure~\ref{fig:beta}.

We extensively examined SSL with different window sizes, $N_f$, for CAPC and all baselines within the flexible RSCNet framework, as shown in Figure~\ref{fig:window_size}. We specifically evaluated window sizes of 10, 20, and 50 for CAPC and CPC because larger $N_f$ values reduce the sequence length $L$, and both CPC and CAPC require $L \geq 3$ to predict future windows effectively. For other methods, we extended our examination to include window sizes of 100 and 200, corresponding to the complete sequence of the CSI in SignFi. Our results indicate that our proposed CAPC, along with Barlow Twins and SimCLR, demonstrated robust performance across different $N_f$ values, with CAPC showing superior accuracy in all tested cases. In contrast, AutoFi and CPC experienced significant performance declines as the number of windows increased.

This suggests that CSI segmentation is a viable approach for SSL in WiFi sensing, benefiting methods beyond our proposed model. Not only did all methods generally perform better with lower $N_f$ values, but, as Figure~\ref{fig:window_size} illustrates, these lower values also simplify the complexity of the encoder $E$ due to smaller input sizes. This reduction enhances efficiency and reduces computational demands, making it particularly suitable for resource-limited edge devices where these models are deployed. Based on these findings, we selected $N_f = 10$ for subsequent experiments, as it generally yielded strong performance across all methods.

\subsection{Autoregressive model selection}

\begin{figure}[h]
    \centering
    \includegraphics[width=1\linewidth]{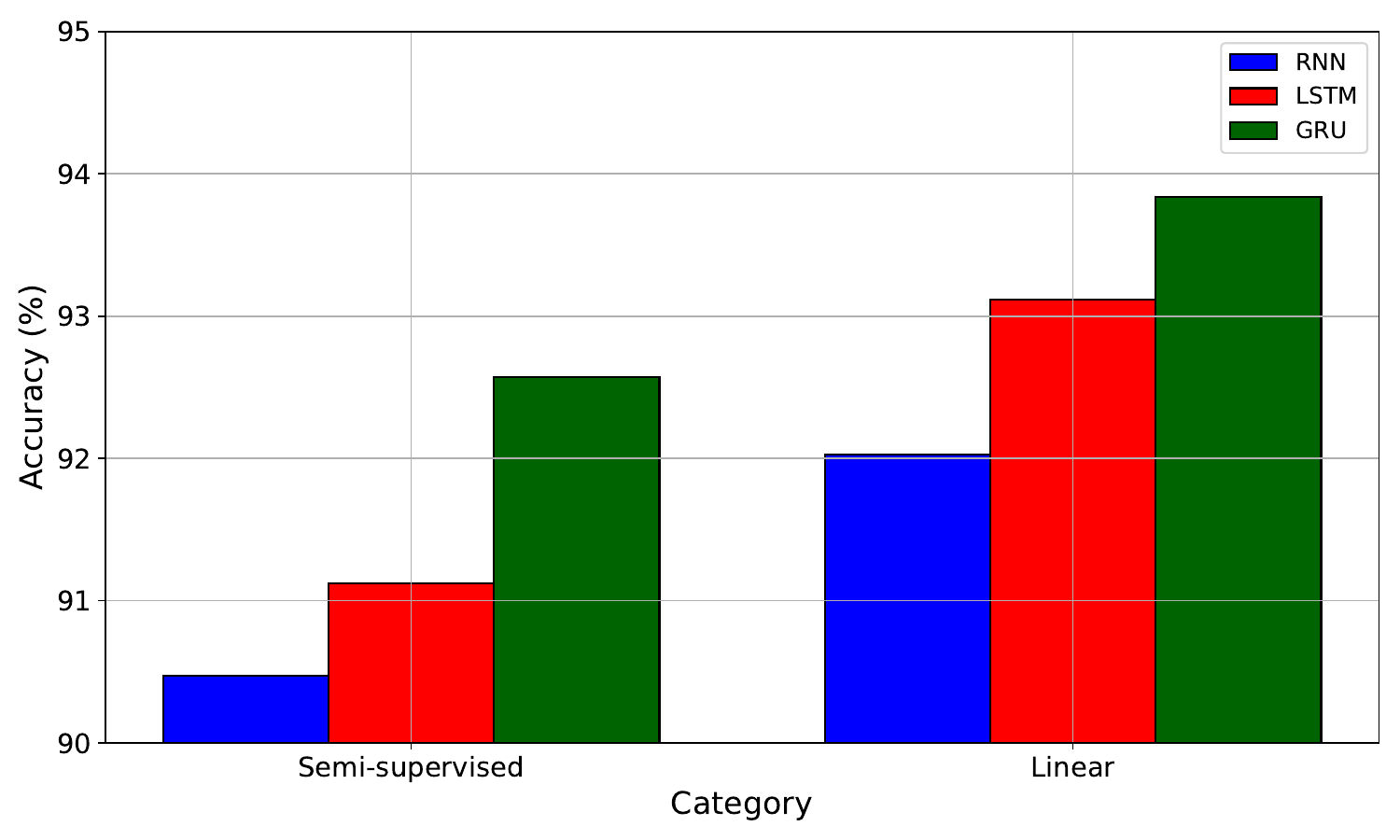}
    \caption{Performance comparison of different autoregressive models (RNN, LSTM, and GRU) for creating context embedding during the self-supervised phase of CAPC. The GRU model shows superior results in both linear and semi-supervised evaluations with 6 shots of SignFi Home dataset, indicating its effectiveness in our framework.}
    \label{fig:autoregressive}
\end{figure}

As shown in Figure \ref{fig:autoregressive} , using GRU as the autoregressive model outperforms both RNN and LSTM for generating context embeddings in the SSL phase of CAPC, achieving the highest accuracy in both semi-supervised and linear evaluations. This advantage is likely due to the GRU's ability to mitigate the vanishing gradient problem. The LSTM also performs well, with 91.12\% accuracy in the semi-supervised setting and 93.11\% in linear evaluations. In contrast, the RNN exhibits the lowest performance, possibly due to its limited memory capacity. Overall, these findings highlight GRU as the optimal choice for our framework.

\subsection{Collapse analysis}
\label{subsec:collapse}

\begin{figure}[h]
    \centering
    \includegraphics[width=0.85\linewidth]{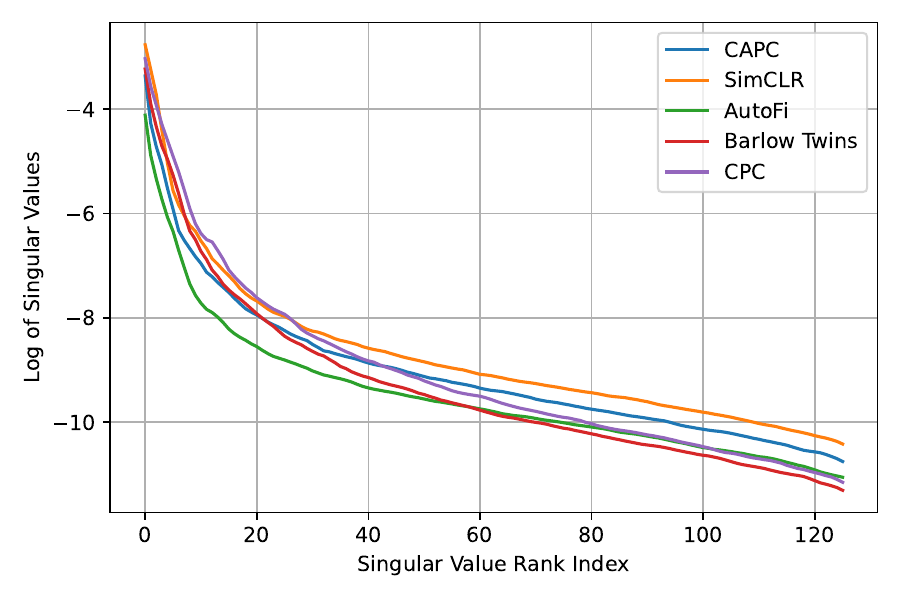}
    \caption{Singular value spectrum of the representation space ($z$) of CAPC compared to baselines on the SignFi Home dataset validation set. Each embedding vector is of size 128. The spectrum displays the singular values of the covariance matrix of these embedding vectors, sorted and plotted on a logarithmic scale. No singular values drop to zero, indicating that none of the methods, including ours, experience dimensional collapse.}
    \label{fig:svd}
\end{figure}

To demonstrate that our method does not suffer from dimensional collapse, we present the singular value spectrum of the embedding space (Figure~\ref{fig:svd}) of the pre-trained encoder of CAPC and other baselines. Complete collapse occurs when all singular values fall to zero, indicating that the representation has become constant and carry no useful information. If only some singular values drop to zero, it suggests partial dimensional collapse, where the encoder has not fully utilized the embedding space \cite{jing2021understanding}. As depicted in Figure~\ref{fig:svd}, no singular values fall to zero for any of the methods, confirming that none, including ours, undergo dimensional collapse.

\subsection{t-SNE visualization}

\begin{figure}[h]
    \centering
    \includegraphics[width=0.85\linewidth]{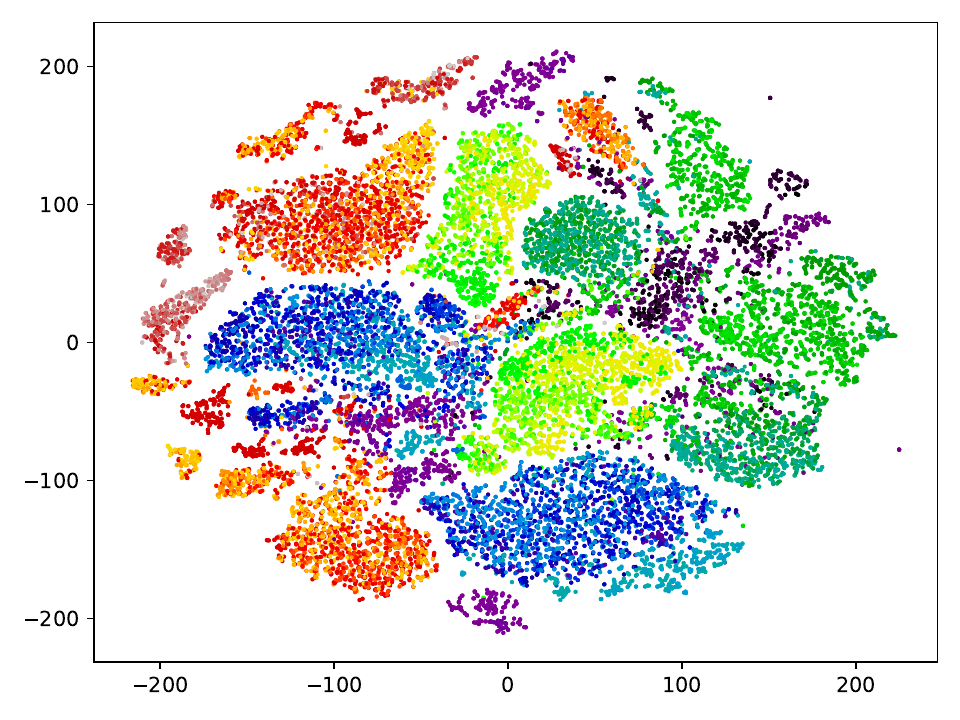}
    \caption{t-SNE visualization of SignFi Home dataset representations, trained using the CAPC SSL method on the SignFi Lab dataset. Each color corresponds to a distinct sign language label.}
    \label{fig:tsne}
\end{figure}

In Figure~\ref{fig:tsne}, we present a t-SNE visualization \cite{van2008visualizing} of the encoded CSI with the proposed CAPC pre-trained encoder, marked by the colored labels in the SignFi Home dataset. This visualization demonstrates the discriminative power of the embeddings for sign language recognition, effectively clustering similar gestures together. Despite the large number of labels and the lack of supervision or predefined labels during training, the encoder effectively segregates the data into distinct clusters. Moreover, the input data originates from an unseen environment, further underscoring the capability of CAPC to not only extract relevant discriminative features for downstream tasks but also to generalize effectively in new environments.

\section{Conclusion}
\label{sec:conclusion}

In this paper, we proposed CAPC, a representation learning framework for CSI data and WiFi sensing, featuring a time-series-specific architecture. Specifically, we employed a hybrid contrastive loss function that combines future prediction and embedding consistency pretext tasks for SSL, ensuring the generated representations are both temporally informative and robust to data distortions inherent in downstream tasks. We also developed a novel augmentation technique to reduce electronic distortions from transceivers and isolate free space propagation effects on the channel. Extensive cross-domain experiments demonstrated that CAPC, with and without dual view augmentation, surpasses baseline SSL methods in downstream HAR and gesture recognition tasks, particularly in few-shot scenarios. Furthermore, using the novel dual view augmentation significantly boosts CAPC's performance. Future works may explore other architectural designs such as the attention mechanism in Transformers \cite{vaswani2017attention} and Fourier Transform based STFNet blocks  \cite{10.1145/3308558.3313426}, integrating other modalities like vision, \cite{yang2024maskfi, deng2022gaitfi, yang2024mm} or CSI streams from multiple WiFi devices \cite{7znf-qp86-20}, as well as considering more complex tasks such as multi-user sensing \cite{huang2024wimans}.


\bibliography{ref}
\bibliographystyle{IEEEtran}

\begin{IEEEbiography}[{\includegraphics[width=1in,height=1.25in,clip,keepaspectratio]{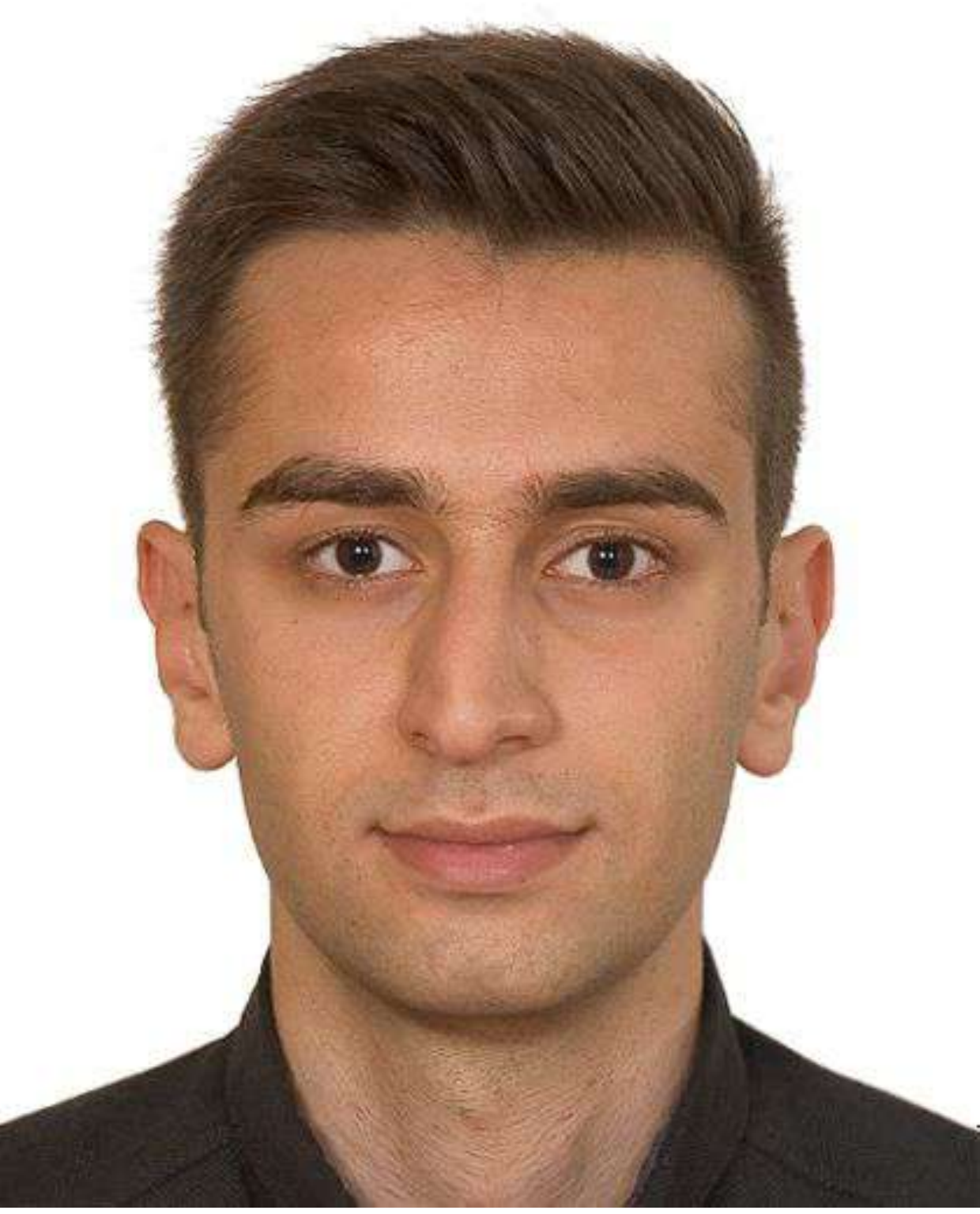}}]{Borna Barahimi } 
\hspace{0.01em} (Graduate Student Member, IEEE) received the B.Sc. degree in computer science from the University of Tehran, Tehran, Iran, in 2022, and the M.Sc. degree in computer science from York University, Toronto, ON, Canada, in 2024. He is currently a Research Assistant with the Next Generation Wireless Networks (NGWN) Lab under the supervision of Dr. Hina Tabassum. His research interests include machine learning and representation learning, with a focus on time-series data and WiFi sensing.
\end{IEEEbiography}

\begin{IEEEbiography}[{\includegraphics[width=1in,height=1.25in, clip,keepaspectratio]{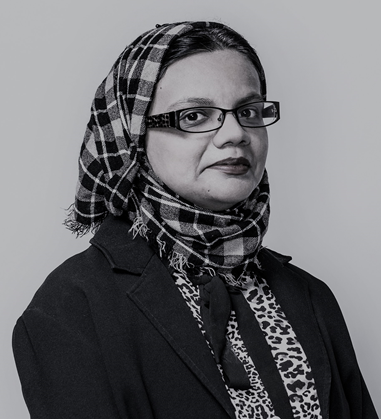}}]{\textbf{Hina Tabassum }} \hspace{0.01em} (Senior Member, IEEE) received the Ph.D. degree from the King Abdullah University of Science and Technology (KAUST). She is currently an Associate Professor with the Lassonde School of Engineering, York University, Canada, where she joined as an Assistant Professor, in 2018. She is currently appointed as the York Research Chair of 5G/6G-enabled mobility and sensing applications (2023 - 2028) and a visiting Professor at University of Toronto (2024-2025). She was a postdoctoral research associate at University of Manitoba, Canada.  She has published over 100 refereed papers in well-reputed IEEE journals, magazines, and conferences. Her research interests include multi-band optical, mm-wave, and THz networks and cutting-edge machine learning solutions for next generation wireless communication and sensing networks. She received the Lassonde Innovation Early-Career Researcher Award in 2023 and the N2Women: Rising Stars in Computer Networking and Communications in 2022. She was listed in the Stanford’s list of the World’s Top $2\%$ Researchers in 2021, 2022, and 2023.  She is the Founding Chair of the Special Interest Group on THz communications in IEEE Communications Society (ComSoc)-Radio Communications Committee (RCC). She served as an Associate Editor for IEEE Communications Letters (2019–2023), IEEE Open Journal of the Communications Society (OJCOMS) (2019–2023), and IEEE Transactions on Green Communications and Networking (TGCN) (2020–2023). Currently, she is also serving as an Area Editor for IEEE OJCOMS and an Associate Editor for IEEE Transactions on Communications, IEEE Transactions on Wireless Communications, and IEEE Communications Surveys and Tutorials. She has been recognized as an Exemplary Editor by the IEEE Communications Letters (2020), IEEE OJCOMS (2023), and IEEE TGCN (2023).  
\end{IEEEbiography}

\begin{IEEEbiography}[{\includegraphics[width=1in,height=2.25in, clip,keepaspectratio]{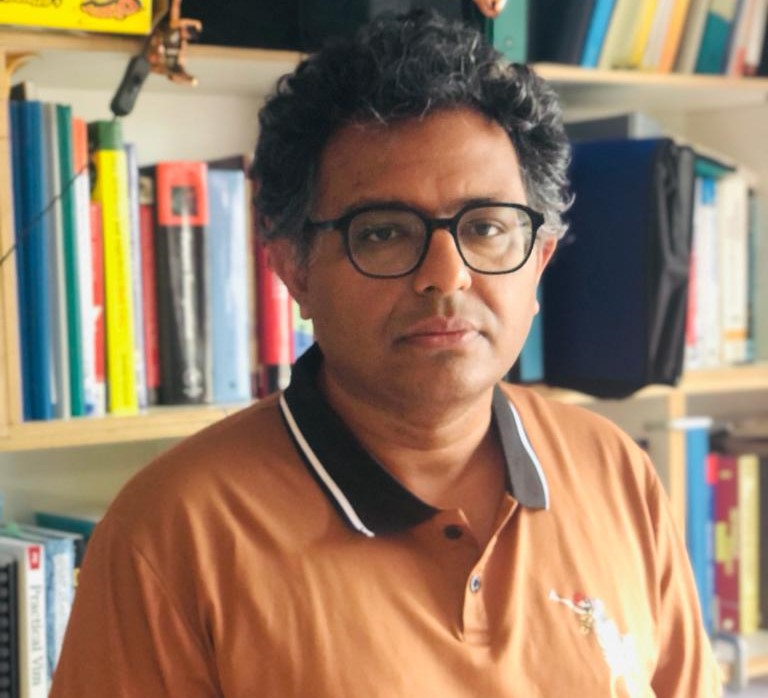}}]{\textbf{Mohammad Omer}} \hspace{0.01em}  obtained his MS and PhD in Engineering from Georgia Institute of Technology. He worked in the advanced radio division of Qualcomm research and several technology startups. Currently, he serves as a Principal Scientist at Cognitive Systems Corp where his work focuses on RF sensing and mapping technologies. He has developed several industry leading solutions for sensing and perception using ambient RF signals, and holds more than 40 US patents.
\end{IEEEbiography}

\begin{IEEEbiography}[{\includegraphics[width=1in,height=1.25in, clip,keepaspectratio]{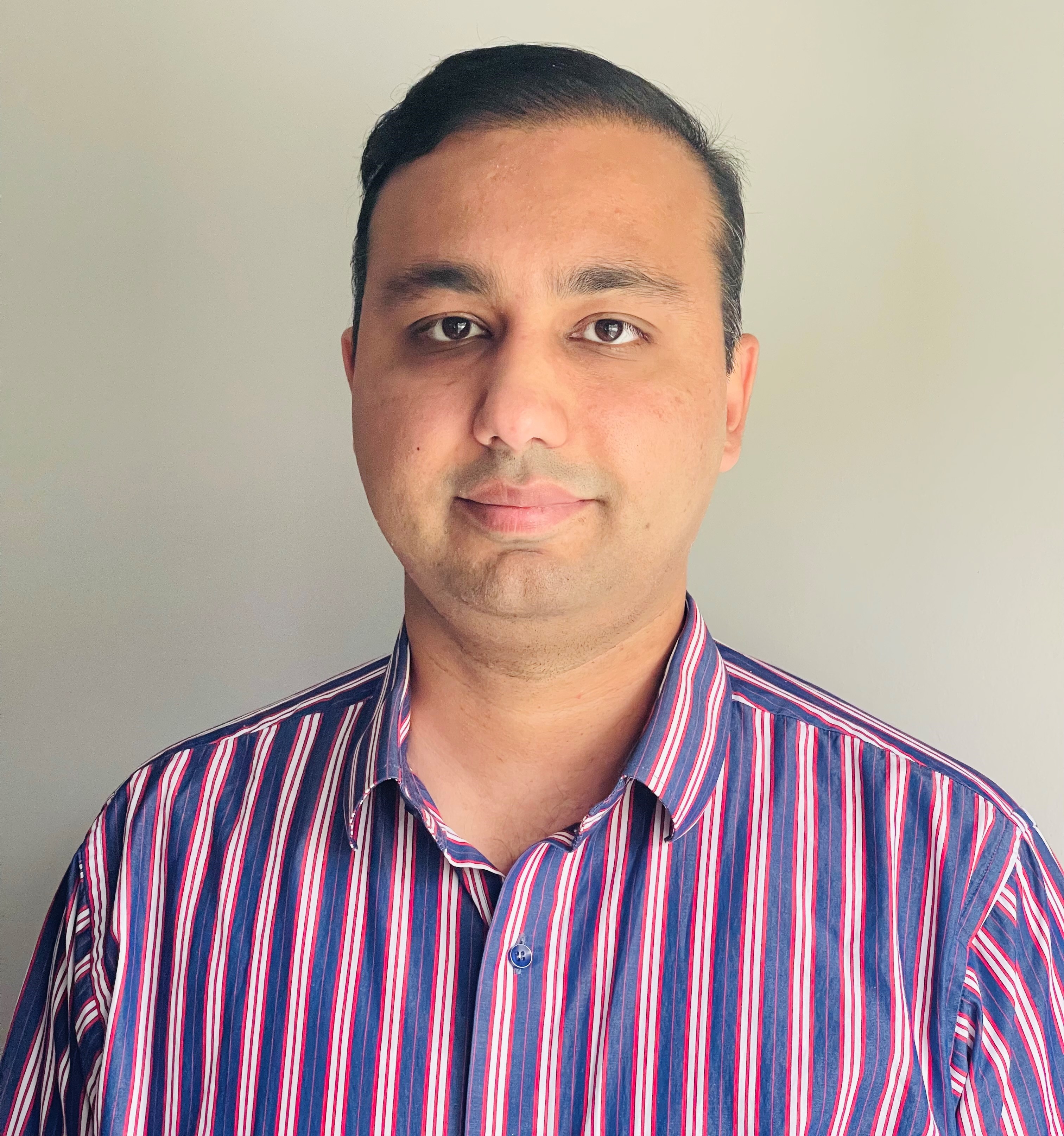}}]{\textbf{Omer Waqar }} \hspace{0.01em} (Senior Member, IEEE) eceived the B.Sc. degree in electrical engineering from the University of Engineering and Technology (UET), Lahore, Pakistan, in 2007, and the Ph.D. degree in electrical and electronic engineering from the University of Leeds, Leeds, U.K., in November 2011. From January 2012 to July 2013, he was a Research Fellow with the Center for Communications Systems Research and 5G Innovation Center, University of Surrey, Guildford, U.K. He worked as an Assistant Professor with UET from August 2013 to June 2018. He worked as a Researcher with the Department of Electrical and Computer Engineering, University of Toronto, Canada, from July 2018 to June 2019. He worked as an Assistant Professor with the Department of Engineering, Thompson Rivers University, Kamloops, BC, Canada, from August 2019 to July 2023. Since August 2023, he has been working as an Assistant Professor with the School of Computing, University of the Fraser Valley, Abbotsford, BC, Canada, and is an Adjunct Faculty with York University, Toronto, ON, Canada. He has authored or coauthored 35+ peer-reviewed articles, including top-tier journals, such as IEEE TRANSACTIONS ON VEHICULAR TECHNOLOGY and IEEE TRANSACTIONS ON MOBILE COMPUTING. His current research interests include reconfigurable intelligent surface-aided communication systems, deep learning for next-generation communication networks, wireless sensing, and resource allocation of wireless networks for several distributed machine learning paradigms. He has secured over \$200K in research grants from the Tri-Council Agency, i.e., NSERC Discovery grant and NSERC Alliance grants. He is currently serving as an Associate Editor for the IEEE OPEN JOURNAL OF THE COMMUNICATIONS SOCIETY and IEEE CANADIAN JOURNAL OF ELECTRICAL AND COMPUTER ENGINEERING.
\end{IEEEbiography}

\end{document}